\documentclass[modern]{aastex62}

\usepackage{graphicx}
\usepackage{xfrac}
\usepackage{bbding}
\usepackage[nointegrals]{wasysym}
\newcommand{\sph}{\text{\fullmoon}}
\newcommand{\obl}{\text{\Ellipse}}

\usepackage{microtype}
\usepackage{url}
\usepackage{amsmath}
\usepackage{mathtools}
\usepackage{esint}
\usepackage{amssymb}
\usepackage{natbib}
\usepackage{multirow}
\usepackage{graphicx}
\usepackage{scalerel}
\usepackage{calc}
\usepackage{etoolbox}
\usepackage{marginnote}
\usepackage{nicefrac}
\usepackage{tabstackengine}
\usepackage{diagbox}
\usepackage[makeroom]{cancel}
\usepackage{mathdots}
\usepackage{bbm}
\usepackage{booktabs}
\usepackage{xspace}
\usepackage{upgreek}
\usepackage[T1]{fontenc} 
\usepackage{textcomp}
\usepackage{ifxetex}
\ifxetex
\usepackage{fontspec}
\defaultfontfeatures{Extension = .otf}
\fi
\usepackage{fontawesome}
\usepackage{listings}
\usepackage{mathtools}
\stackMath

\definecolor{editcolor}{rgb}{0,0,0}

\newcommand{\starry}{\textsf{starry}\xspace}

\renewcommand{\eqref}[1]{\ref{eq:#1}}

\definecolor{linkcolor}{rgb}{0.1216,0.4667,0.7059}
\newcommand{\codeicon}{{\color{linkcolor}\faFileCodeO}}
\newcommand{\prooficon}{{\color{linkcolor}\faPencilSquareO}}
\newcommand{\animicon}{{\color{linkcolor}\faPlayCircle}}
\newcommand{\codelink}[1]{\href{https://github.com/rodluger/starry/blob/v0.2.2/tex/figures/#1.py}{\codeicon}\,\,}
\newcommand{\animlink}[1]{\href{https://github.com/rodluger/starry/blob/v0.2.2/tex/figures/#1.gif}{\animicon}\,\,}
\newcommand{\prooflink}[1]{\href{https://github.com/rodluger/starry/blob/v0.2.2/docs/proofs/#1.ipynb}{\raisebox{-0.1em}{\prooficon}}}

\newtagform{eqtag}[]{(}{)}
\newcommand{\currentlabel}{None}

\newenvironment{proof*}[1]{%
\ifstrempty{#1}{%
\renewtagform{eqtag}[]{\raisebox{-0.1em}{{\color{red}\faPencilSquareO}}\,(}{)}%
}{%
\renewtagform{eqtag}[]{\prooflink{#1}\,(}{)}%
}%
\usetagform{eqtag}%
\renewcommand{\currentlabel}{#1}
\equation%
}{%
\endequation%
\renewtagform{eqtag}[]{(}{)}%
\usetagform{eqtag}%
\message{<<<\currentlabel: \theequation>>>}
}

\newcommand{\dd}{\ensuremath{ \mathrm{d}}}

\newcommand{\bvec}[1]{{\ensuremath{\mathbf{#1}}}}

\newcommand{\x}{\ensuremath{\mbox{$x$}}}
\newcommand{\y}{\ensuremath{\mbox{$y$}}}
\newcommand{\z}{\ensuremath{\mbox{$z$}}}
\newcommand{\xhat}{\ensuremath{\mathbf{\hat{x}}}\xspace}
\newcommand{\yhat}{\ensuremath{\mathbf{\hat{y}}}\xspace}
\newcommand{\zhat}{\ensuremath{\mathbf{\hat{z}}}\xspace}
\DeclareMathAlphabet\mathbfcal{OMS}{cmsy}{b}{n}

\DeclarePairedDelimiter\floor{\lfloor}{\rfloor}
\definecolor{dim}{rgb}{0.8,0.8,0.8}
\newcommand{\gbasis}{\ensuremath{\tilde{\bvec{g}}}}

\newcommand{\gbasisn}{\ensuremath{\tilde{g}_n}}

\definecolor{codegreen}{rgb}{0,0.6,0}
\definecolor{codegray}{rgb}{0.5,0.5,0.5}
\definecolor{codepurple}{rgb}{0.58,0,0.82}
\definecolor{backcolour}{rgb}{0.95,0.95,0.95}
\lstdefinestyle{mystyle}{
    backgroundcolor=\color{backcolour},
    commentstyle=\color{codegreen},
    keywordstyle=\color{magenta},
    numberstyle=\tiny\color{codegray},
    stringstyle=\color{codepurple},
    basicstyle=\small\ttfamily,
    breakatwhitespace=false,
    breaklines=true,
    captionpos=b,
    keepspaces=true,
    numbers=left,
    numbersep=5pt,
    showspaces=false,
    showstringspaces=false,
    showtabs=false,
    tabsize=2,
    aboveskip=1em,
    belowskip=1em,
    keywords=[2]{map},
    keywordstyle=[2]{\color{black!80!black}},
    upquote=true
}
\makeatletter
\def\Ddots{\mathinner{\mkern1mu\raise\p@
\vbox{\kern7\p@\hbox{.}}\mkern2mu
\raise4\p@\hbox{.}\mkern2mu\raise7\p@\hbox{.}\mkern1mu}}
\makeatother

\renewcommand\quad{\hskip\fontdimen3\font}

\graphicspath{{./}{figures/}}

\submitjournal{ApJ}

\shorttitle{Gravity-darkened transits}
\shortauthors{Dholakia et al.}

\usepackage{lineno}

\begin{document}

\title{Efficient and precise transit light curves for rapidly-rotating, oblate stars}

\correspondingauthor{Shashank Dholakia}
\email{dholakia.shashank@berkeley.edu}

\author[0000-0001-9145-8444]{Shashank Dholakia}
\affil{Department~of~Astronomy, University of California, Berkeley, CA}

\author[0000-0002-0296-3826]{Rodrigo Luger}
\affil{Center~for~Computational~Astrophysics, Flatiron~Institute, New~York, NY}

\author[0000-0001-6263-4437]{Shishir Dholakia}
\affil{Department~of~Astronomy, University of California, Berkeley, CA}



\begin{abstract}

We derive solutions to transit light curves of exoplanets orbiting rapidly-rotating stars. These stars exhibit significant oblateness and gravity darkening, a phenomenon where the poles of the star have a higher temperature and luminosity than the equator. Light curves for exoplanets transiting these stars can exhibit deviations from those of slowly-rotating stars, even displaying significantly asymmetric transits depending on the system's spin-orbit angle. As such, these phenomena can be used as a protractor to measure the spin-orbit alignment of the system. In this paper, we introduce a novel semi-analytic method for generating model light curves for gravity-darkened and oblate stars with transiting exoplanets. We implement the model within the code package \starry and demonstrate several orders of magnitude improvement in speed and precision over existing methods. We test the model on a TESS light curve of WASP-33, whose host star displays rapid rotation ($v \sin i_* = 86.4$ km/s). We subtract the host's $\delta$-Scuti pulsations from the light curve, finding an asymmetric transit characteristic of gravity darkening. We find the projected spin orbit angle is consistent with Doppler tomography and constrain the true spin-orbit angle of the system as $\varphi=108.3^{+19.0}_{-15.4}$~$^{\circ}$. We demonstrate the method's uses in constraining spin-orbit inclinations of such systems photometrically with posterior inference. Lastly, we note the use of such a method for inferring the dynamical history of thousands of such systems discovered by TESS. \href{https://github.com/rodluger/starry}{\color{linkcolor}\faGithub} \href{https://rodluger.github.io/starry}{\color{linkcolor}\faBook}
\href{https://github.com/shashankdholakia/gravity-dark}{\color{linkcolor}\faBarChart}
\end{abstract}



\section{Introduction} 
\label{sec:intro}

The exoplanet transit method has led to the discovery of thousands of exoplanet candidates since the first observed transits of the HD 209458 system by \citet{charbonneau2000} and \citet{henry2000}. The utility of the transit method to discover and characterize exoplanet systems was largely based on the development of fast and precise photometric transit models---functions describing the change in the observed stellar flux over the course of a transit given a certain set of orbital and planetary parameters. These models can be fit to a stellar light curve and used to infer the best-fitting planetary and orbital parameters. 
One of the most widely adopted transit models is that of \citet{mandel2002}, who found an analytic (i.e., closed form) solution to the integral describing the flux blocked by a planet transiting a star with quadratic limb-darkening. This analytic model permitted the generation of fast model light curves, allowing thousands of exoplanet candidates to have their best-fitting parameters measured, including orbital period, planet-star radius ratio, and inclination.

Despite its applicability to thousands of transiting exoplanets, the \citet{mandel2002} model makes some assumptions that break down for certain systems. For instance, it assumes that the intensity profile across the stellar disk can be modeled using a single quadratic function in the radial parameter $\mu$ to capture the effect of limb darkening. Secondly, it assumes that both the planet and star can be modeled as circles in projection. Both of these assumptions break down in the case of planets transiting rapidly-rotating stars. 

Early-type main sequence stars with temperatures above about 6200K---a limit known as the \emph{Kraft break} \citep{kraft1967}---exhibit an abrupt increase in mean rotational velocities. This increase is due to the fact that these stars lack the strong convective magnetic dynamos present in cooler stars that transfer angular momentum out of the star through stellar winds \citep{parker1958,royer2009}. These rapidly-rotating stars can be significantly oblate due to the centrifugal force pulling the equator outwards. 
Additionally, pre-main sequence stars and some young main sequence M dwarfs have been shown to display very rapid rotation \citep{rebull2016, rebull2017, rebull2018, rebull2020, stauffer2018, gilhool2018} and are therefore also likely to be oblate. Other kinds of stars such as binary stars and exoplanet hosts may also exhibit deviations from spherical geometry due to tidal forces in addition to rotation \citep{prsa2018, welsh2010}.

All these stars can also exhibit gravity darkening, an effect where the temperature of the star at the poles is higher than at the equator due to differences in density across the star. For rapidly rotating stars, this is due to the centrifugal force, which results in a lower density at the equator relative to the poles. This temperature difference manifests as a latitude-dependent surface brightness variation, where the equator can be significantly cooler---and therefore less luminous---than the two poles. 

Modeling transits without incorporating stellar rotation effects can lead to systematic errors in planet parameters \citep{barnes2009}. However, far more importantly, these rotation effects can serve as a protractor by which to measure the spin-orbit inclination of an exoplanet system. Spin-orbit angles can be major clues to the formation history of hot-Jupiters and other exoplanets, distinguishing between scenarios such as tidal migration, planet-planet scattering and disc migration. 

Various methods of measuring spin-orbit angles exist and have been used in the literature. All of them utilize some aspect of stellar rotation that breaks spherical symmetry to measure the planet's orbit. By far the most widely-used method is Doppler tomography, for instance used recently by \citet{borsa2021} on the WASP-33 system. Doppler tomography involves measurement of a transit spectroscopically; due to the redshift and blueshift of different sides of the rotating star, the stellar absorption lines are broadened. When a planet transits and occults either blueshifted or redshifted stellar light, it leaves a characteristic ``bump'' on the line profile. Using the path of the bump, the projected spin-orbit angle can be measured with great precision \citep{johnson2015}. Another method that is sometimes used to constrain projected spin-orbit obliquities is the starspot method (\citealt{nutzman2011}, and more recently \citealt{dai2018}). This method involves the determination of the stellar rotation period from starspot-induced variability in the lightcurve. Then, starspot crossings in transit are analyzed for signs that the planet is transiting the same starspot, which would indicate an aligned orbit. Yet another method of constraining spin-orbit information is through asteroseismology, which can provide the stellar inclination angle $i_*$ through comparison of the relative power of nonradial and radial modes \citep{benomar2014}. 

All of the above methods constrain mainly one component of the true spin-orbit angle: in the case of Doppler tomography and the starspot method, the projected spin-orbit angle and in the case of asteroseismology the mutual inclination between the planet and the star. However, taking into account gravity darkening and oblateness in a transit model has the ability to provide true spin-orbit angles from a single measurement, as done in e.g. \citep{ahlers2020a}. This makes the method attractive for use of larger datasets such as TESS and Kepler.

In this paper, we develop an efficient, semi-analytic solution to the transits of planets across oblate stars and implement this in the code package \starry, demonstrating two to three orders of magnitude of speedup over existing methods and several orders of magnitude more precision. 
This paper is organized as follows: In \S\ref{sec:gravdark}, we describe and derive our model for the gravity-darkened stellar intensity profile. In \S\ref{sec:integration}, we devise a technique to integrate the visible flux for an oblate star transited by a spherical planet. In \S\ref{sec:performance}, we benchmark the the model we present against existing models and find significant improvements in accuracy and efficiency. In \S\ref{sec:wasp33}, we apply our model to WASP-33b and find that the oblate, gravity-darkened model is a significant improvement over the spherical host star models that have been applied to it in the past. Finally, in \S\ref{sec:conclusion} we enumerate where this model may be applied and to what effect, as well as outline future improvements and implementations for this model. We leave most of the details of the integration and the math to the Appendix.

\section{Modeling Gravity-darkening and Oblateness} 
\label{sec:gravdark}

\begin{figure}[p!]
\gridline{\fig{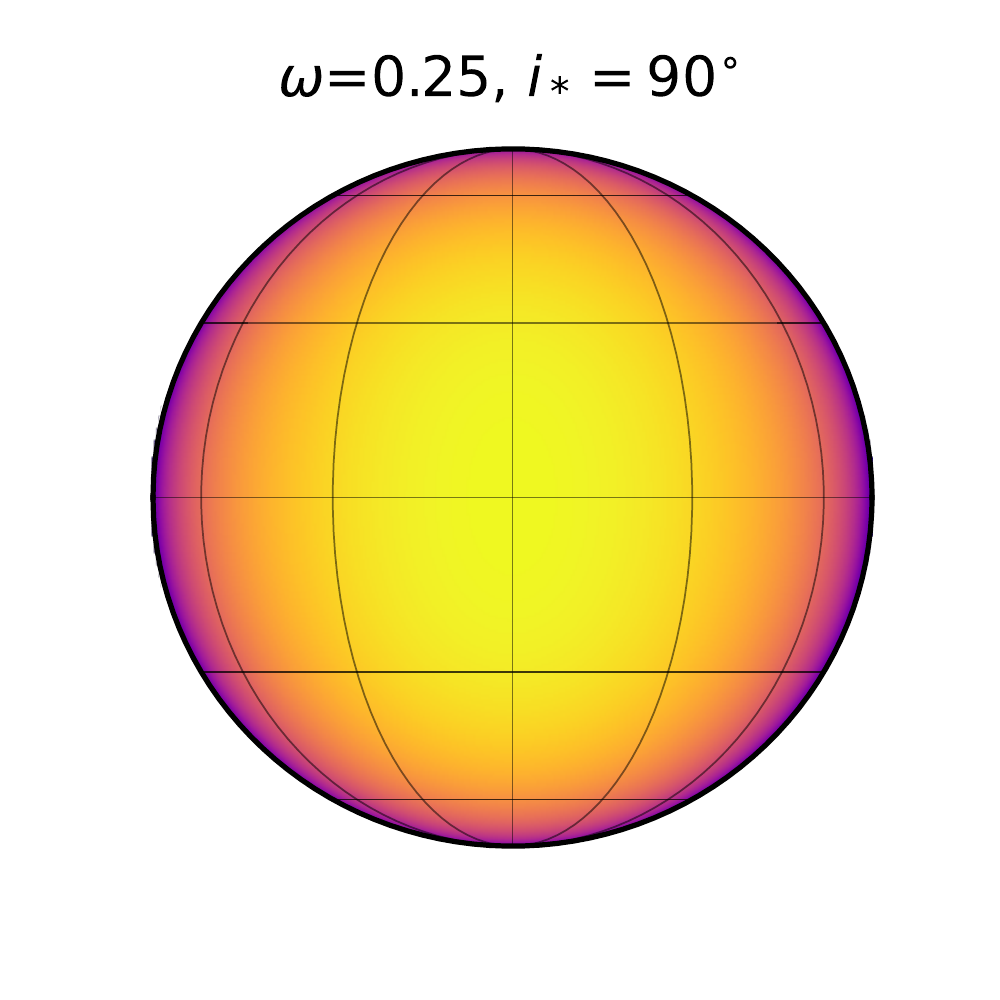}{0.3\textwidth}{}
          \fig{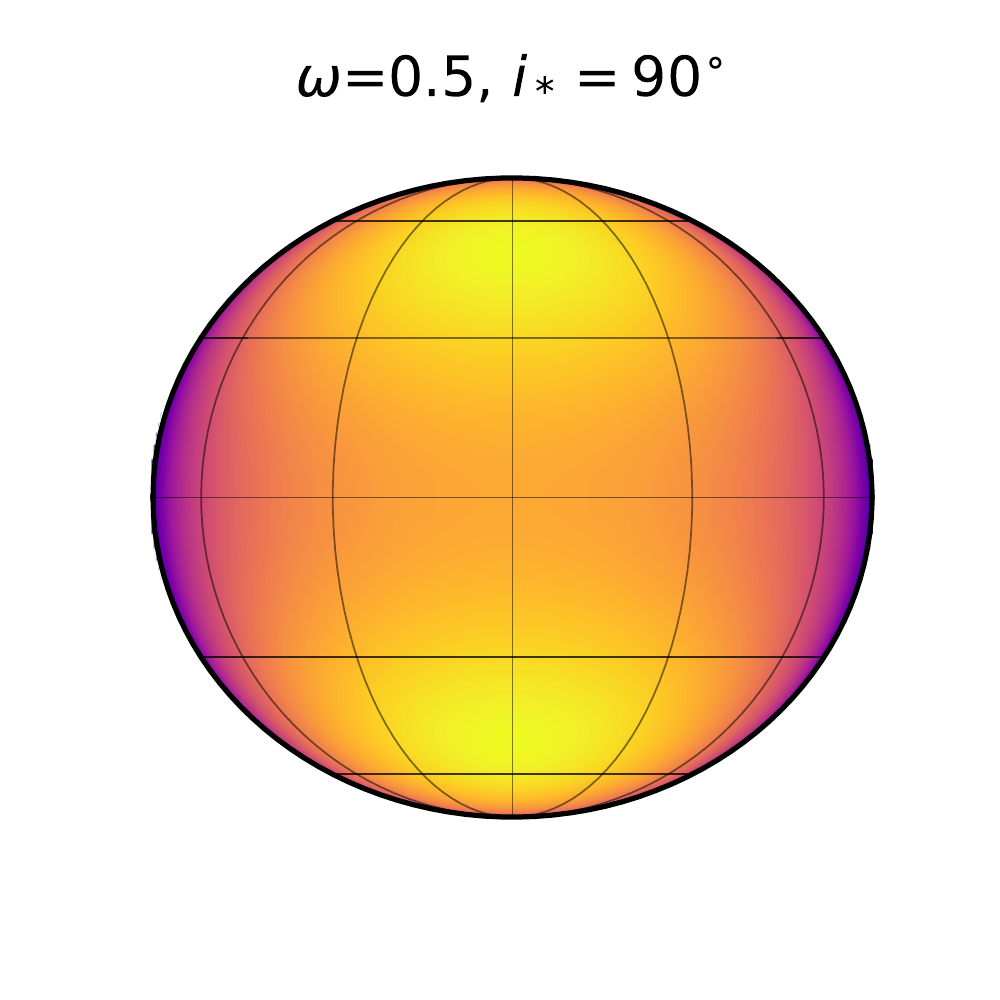}{0.3\textwidth}{}
          \fig{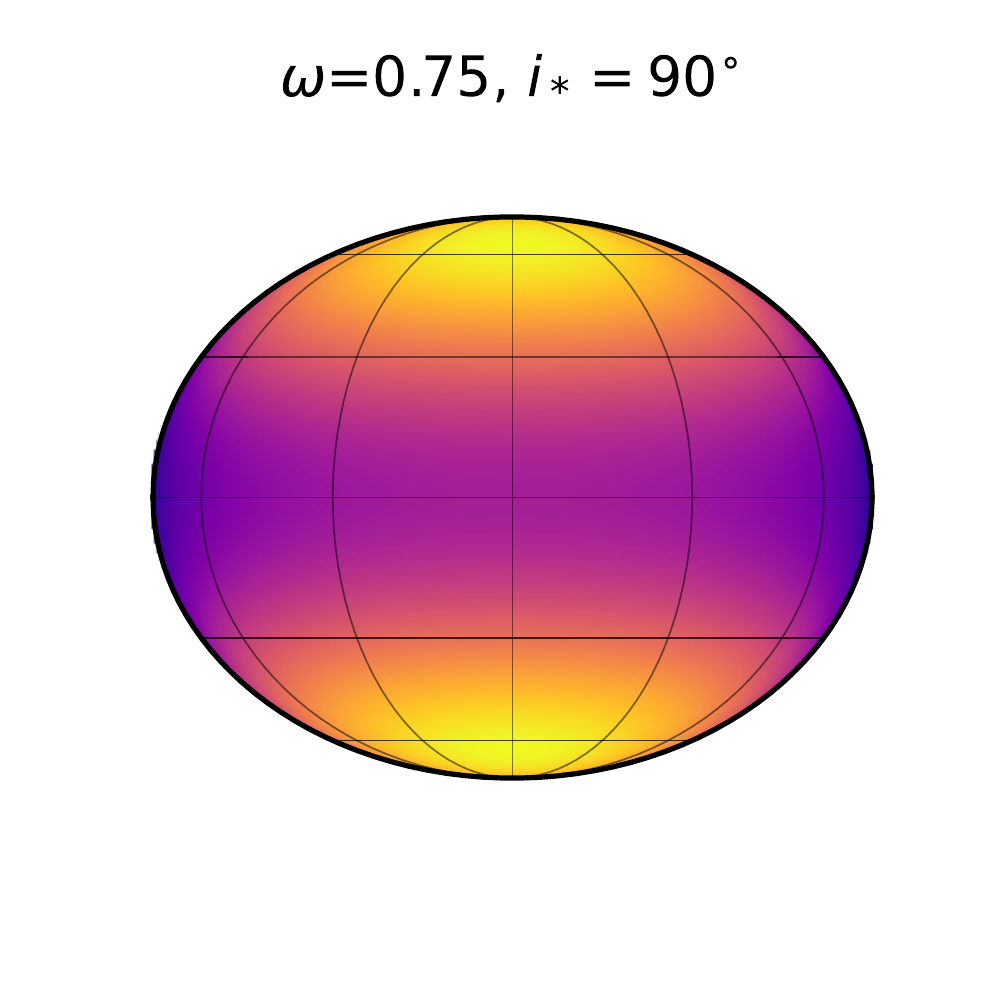}{0.3\textwidth}{}}
\gridline{\fig{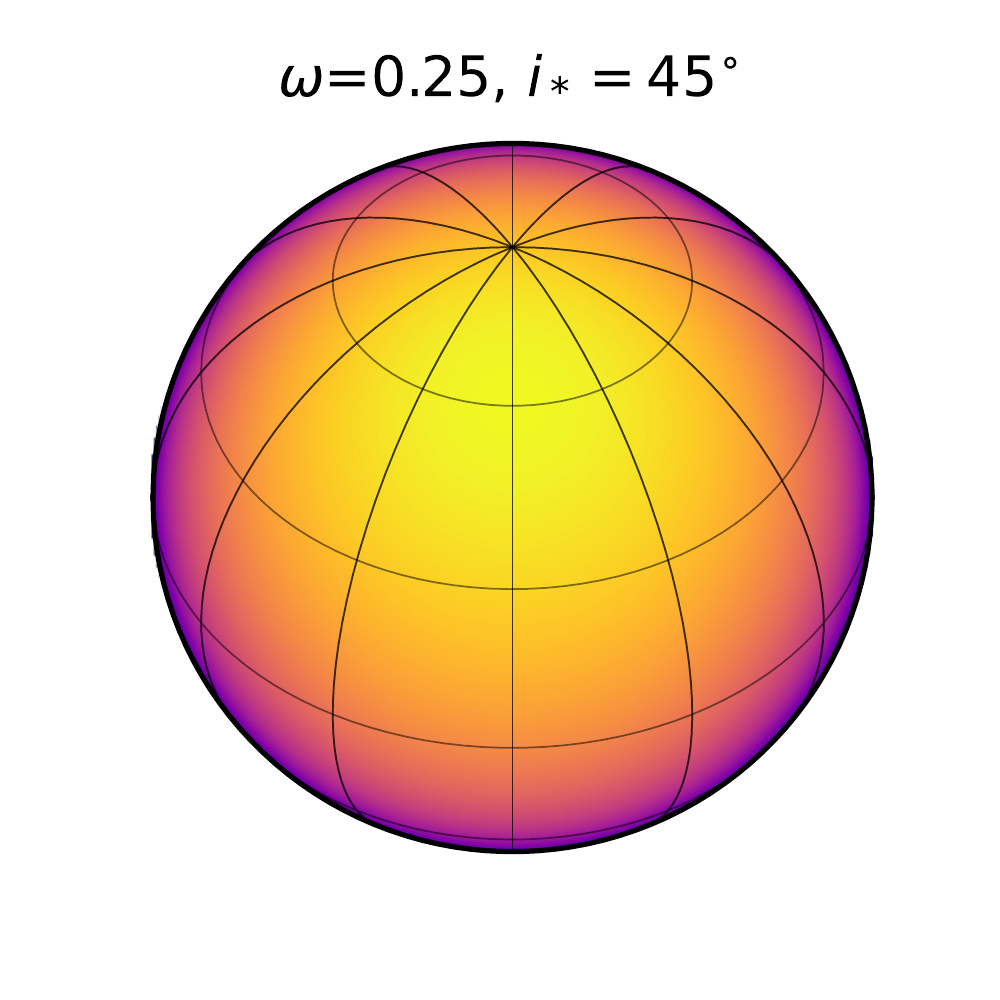}{0.3\textwidth}{}
          \fig{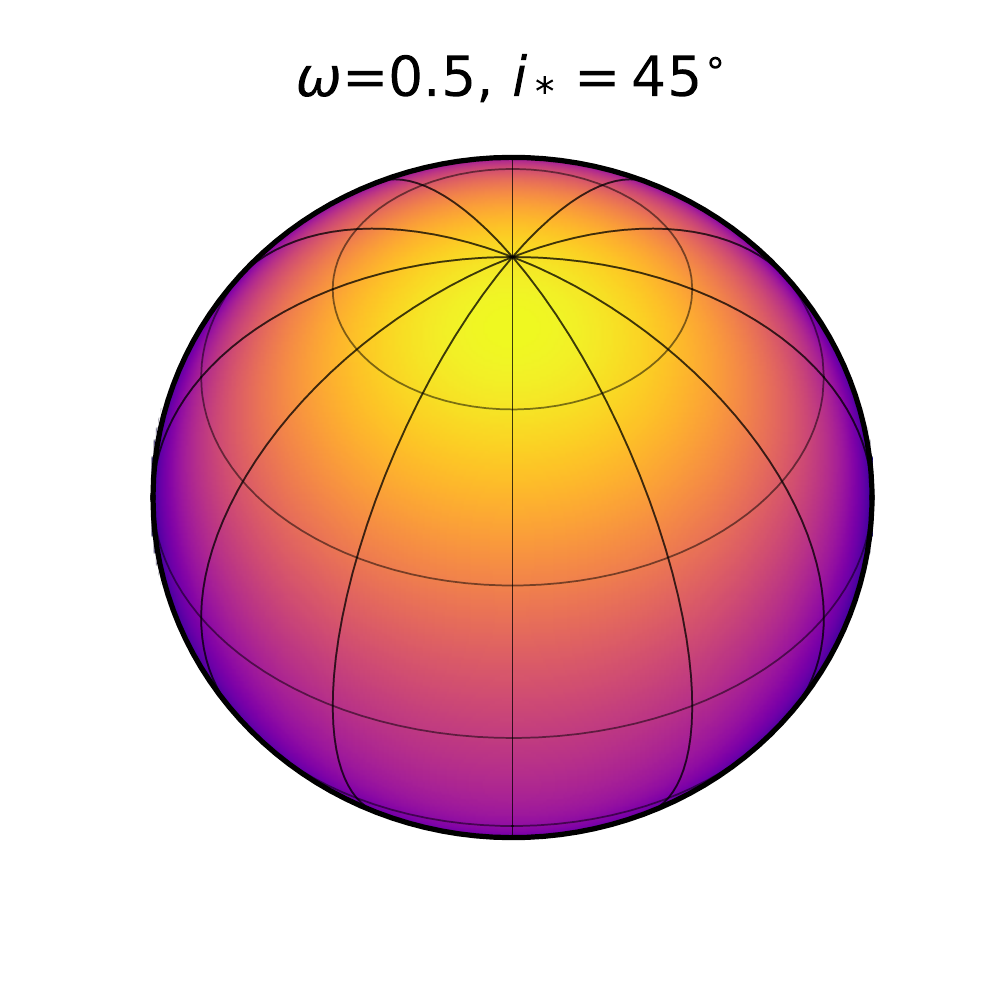}{0.3\textwidth}{}
          \fig{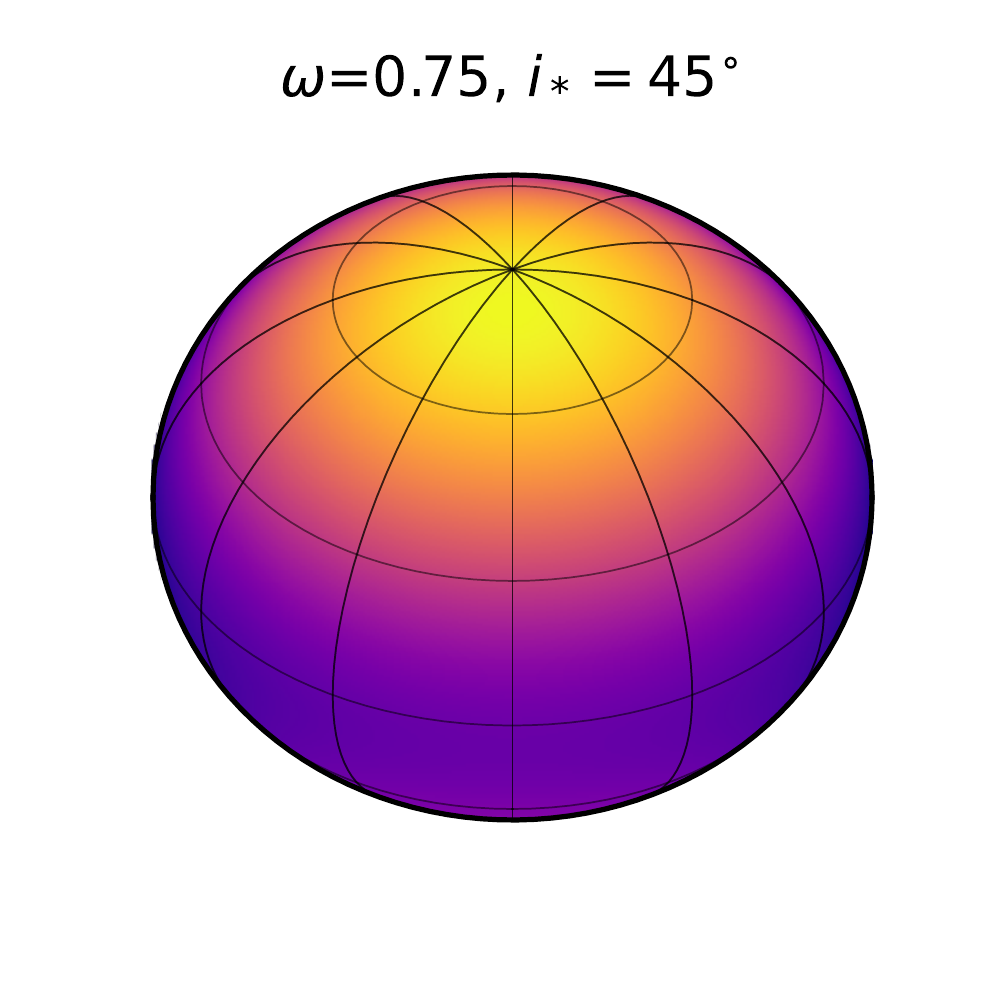}{0.3\textwidth}{}}
\gridline{\fig{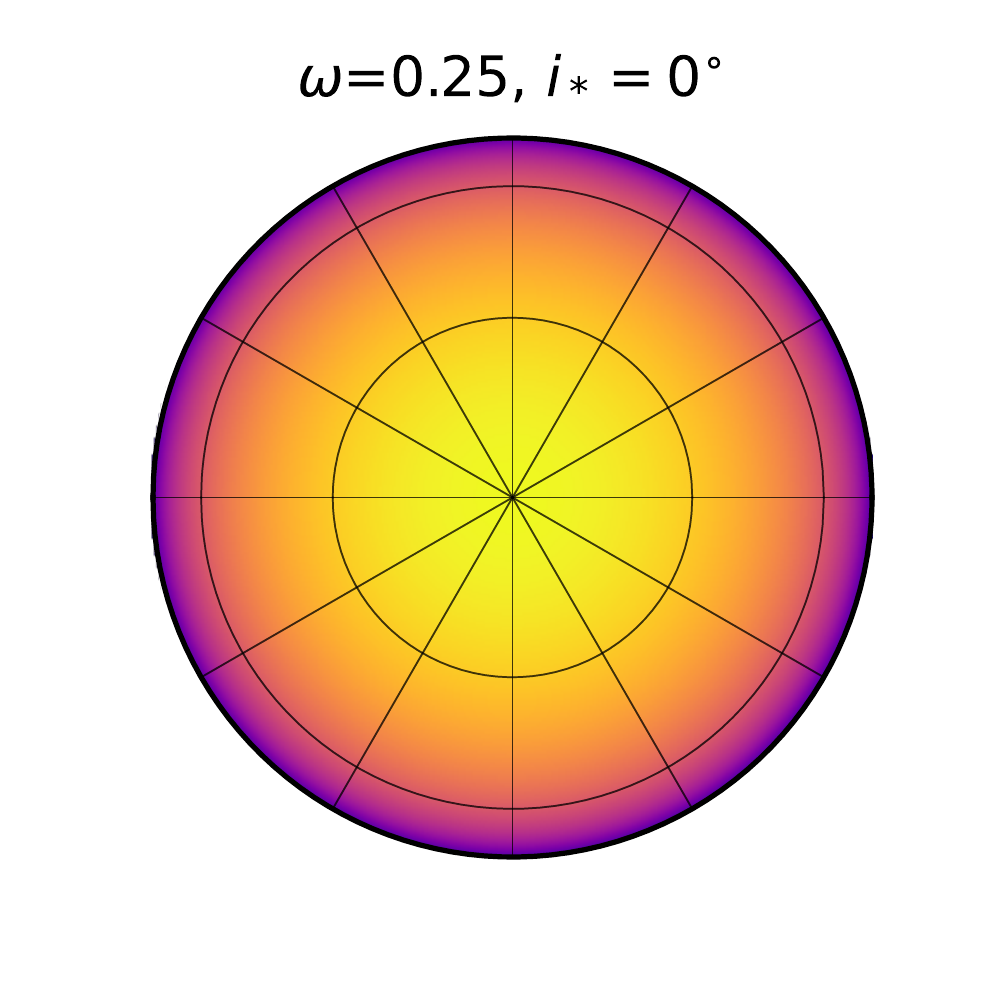}{0.3\textwidth}{}
          \fig{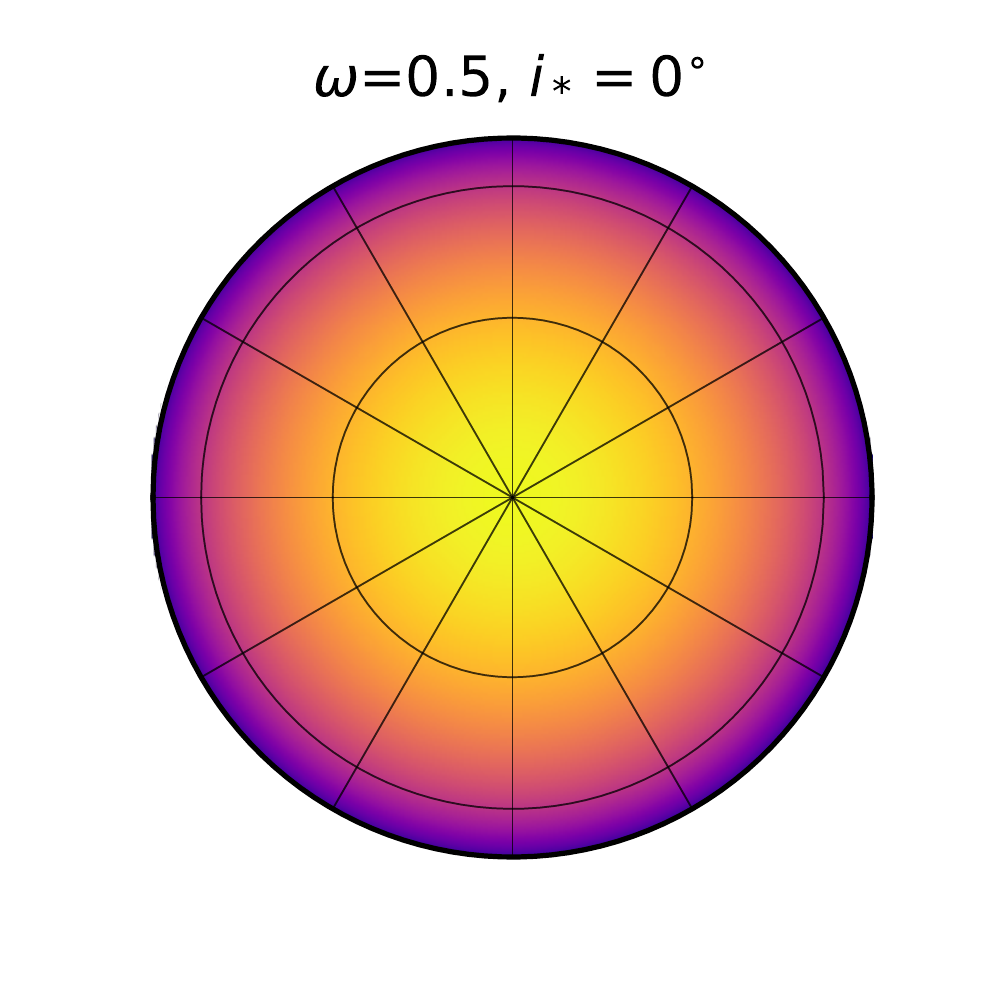}{0.3\textwidth}{}
          \fig{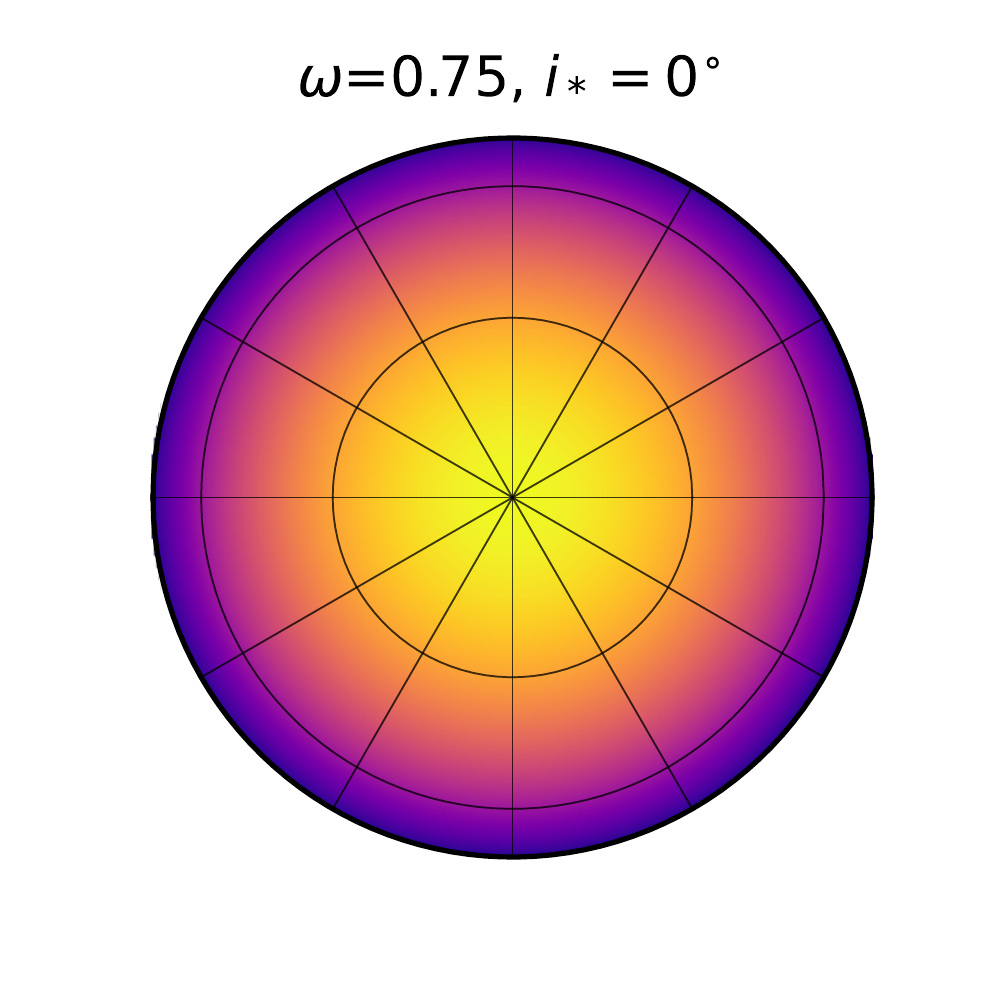}{0.3\textwidth}{}}
\caption{Rapidly-rotating stars at a variety of rotation rates (slowest to fastest from left to right) and orientations (from top to bottom: equator-on, 45$^{\circ}$, pole-on). Parameters such as limb darkening ($u_1=0.2$, $u_2=0.2$), polar temperature ($8500$K), gravity darkening coefficient $\beta = 0.23$ are intended to represent an standard A-type star and the wavelength is taken as 800nm, about the center of the TESS bandpass.}

\label{fig:rrstar}
\end{figure}

\subsection{Current models of gravity-darkened, oblate transits}

Gravity darkening and oblateness have been used to constrain the spin-orbit angles of nine exoplanets to date (to our knowledge in the literature): Kelt-9b \citep{ahlers2020}, MASCARA-4b \citep{ahlers2020a}, WASP-189 \citep{lendl2020}, Hat-P-69b, Hat-P-70b \citep{zhou2019b}, HATS-70b \citep{zhou2019a} Kepler13Ab, Hat-P-7b \citep{masuda2015}, and KOI-189 \citep{barnes2011}. All these measurements utilize the \citet{barnes2009} transit model for oblate and gravity-darkened stars. The \citet{barnes2009} model represents the intensity profile of a rapidly rotating star using the Von Zeipel Theorem on the surface of an oblate MacLaurin spheroid. The model then represents the flux in transit as a 2-D surface integral at each time interval of the light curve. 

A variety of implementations of the \citet{barnes2009} transit model use numerical techniques to solve the 2-D flux integral presented in Equation~(7) of that paper, including \textsf{simuTrans} \citep{armstrong2019} and \textsf{PyTransit} \citep{parviainen2015}. Both of these codes discretize the surface of the star into a grid with a designated resolution and solve for the flux blocked in every point of the transit. These numerical integration are extensible enough to allow a range of models, such as the Von Zeipel law \citep{vonzeipel1924} or the more recent model in \citet{espinosa2011}. However, such techniques are computationally intensive and involve a fundamental tradeoff between precision and speed. A similar but separate technique of modeling the oblateness of planets transiting spherical stars has also been explored with numerical methods in the literature, including by \citet{barnes2003} and \citet{cardao2020} using a Monte Carlo based approach. Codes to model binary systems, such as \textsf{ellec} \citep{maxted2016} and \textsf{PHOEBE} \citep{prsa2016} also model stars as ellipsoids and incorporate gravity-darkened surface maps. Both these codes also use discretization to integrate the flux numerically; \textsf{ellec} models the surface map using a 2-D non-rectangular grid while \textsf{PHOEBE} uses a triangle mesh grid. The ubiquity of such computationally expensive methods to model transits for oblate celestial bodies warrants an attempt at a more efficient and precise solution. 

\subsection{Gravity-darkened surface maps}

Gravity darkening on the surface of a radiative rapidly rotating star is usually taken to be proportional to some power of the local effective surface gravity $g_{\mathrm{eff}}$. This is called the Von Zeipel Law \citep{vonzeipel1924}:

\begin{linenomath}
\begin{equation}
\label{eq:T}
T(x_0, y_0, z_0) = T_{pole} \frac{g(x_0, y_0, z_0)^{\beta}}{g_{pole}^{\beta}}
\end{equation} 
\end{linenomath}
where $T_{pole}$ is the temperature at the pole, $\beta$ is the gravity darkening coefficient, a scaling parameter, and $T$ and $g$ are the temperature and magnitude of the stellar surface gravity at a given point $(x_0, y_0, z_0)$ on the stellar surface. Here, $x_0$, $y_0$ and $z_0$ are the right-handed Cartesian coordinates in a frame where the star is at the origin and the star's rotation axis is aligned with the $+y_0$-axis.%
 
Following \citet{espinosa2011}, we express the angular velocity of the star $\Omega$ as a dimensionless variable $\omega$: 
\begin{linenomath}
\begin{equation} \label{eq:omega}
\omega \equiv \Omega \sqrt{\frac{R_{eq}^3}{GM}}
\end{equation}
\end{linenomath}  
where $R_{eq}$ is the stellar equatorial radius, $G$ is the gravitational constant, and $M$ is the stellar mass.
The surface gravity vector at any point on the star can then be written as a function of this dimensionless rotation rate \citep{barnes2009}:
\begin{linenomath} \begin{equation}
\bvec{g}(x_0, y_0, z_0) = \left[
\begin{matrix}
- \frac{G M x_0}{\left(x_0^{2} + y_0^{2} + z_0^{2}\right)^{\frac{3}{2}}} + \frac{G M \omega^{2} x_0}{R_{eq}^{3}}\\

- \frac{G M y_0}{\left(x_0^{2} + y_0^{2} + z_0^{2}\right)^{\frac{3}{2}}}\\

- \frac{G M z_0}{\left(x_0^{2} + y_0^{2} + z_0^{2}\right)^{\frac{3}{2}}} + \frac{G M \omega^{2} z_0}{R_{eq}^{3}}

\end{matrix}
\right]
\end{equation} \end{linenomath} 
If we represent the star as an oblate spheroid, points on the surface satisfy the relation
\begin{linenomath} \begin{equation}
x_0^2 + \left(\frac{y_0}{1 - f}\right)^2 + z_0^2 = R_{eq}^2,
\label{eq:ellipsoid}
\end{equation} \end{linenomath} 
where
\begin{linenomath}  \begin{align}
    \label{eq:fdef}
    f \equiv \frac{R_{eq} - R_{pole}}{R_{eq}}
    \quad,
    \quad\quad\quad\quad\quad\quad
    0 \leq f < 1
\end{align}  \end{linenomath}
is the normalized difference between the polar radius and the equatorial radius, often called the \emph{oblateness} or \emph{flattening} of the ellipsoid.
Determining the value of the oblateness $f$ is a historied problem \citep{maclaurin1742, chandrasekhar1967} with various approximations, summarized in \citet{essen2004}. For simplicity, here we adopt the approximation used by \citet{masuda2015},
\begin{linenomath} \begin{equation}
    f = \frac{\Omega R_{eq}^3}{2GM}
    \quad.
\end{equation} \end{linenomath} 
Replacing $\Omega$ with the dimensionless quantity $\omega$ from Equation~(\ref{eq:omega}), we obtain
\begin{linenomath} \begin{equation}
    f = 1 - \frac{2}{(\omega^2 + 2)}
    \quad.
\end{equation} \end{linenomath} 

Next, using the fact that at the pole ($x_0 = z_0 = 0$, $y_0 =(1 - f) R_{eq}$) the magnitude of the surface gravity vector is
\begin{linenomath} \begin{equation}
    \bvec{g}_{\mathrm{pole}} = \frac{G M}{R_{eq}^{2} (1 - f)^{2}}
    \quad,
\end{equation} \end{linenomath} 
we can re-write Equation~(\ref{eq:T}) as
\begin{linenomath} \begin{equation}
T(x_0, y_0, z_0) = T_{pole} \left(\frac{(1 - f)^{2} \sqrt{R_{eq}^{6} y_0^{2} + \left(R_{eq}^{3} - \omega^{2} \left(x_0^{2} + y_0^{2} + z_0^{2}\right)^{\frac{3}{2}}\right)^{2} \left(x_0^{2} + z_0^{2}\right)}}{R_{eq} \left(x_0^{2} + y_0^{2} + z_0^{2}\right)^{\frac{3}{2}}}\right)^{\beta}
\quad.
\end{equation} \end{linenomath} 
We see that due to the use of the dimensionless quantity $\omega$, any dependence on $G$ or $M$ cancels, resulting in a gravity darkening profile that is a pure function of spatial variables. 

At this point, it is convenient to normalize all spatial quantities to the equatorial radius of the star, i.e., to adopt spatial units such that $R_{eq} = 1$. Further, since the magnitude of the gravity vector is azimuthally symmetric, the temperature can be written exclusively in terms of the $y_0$ coordinate of points on the stellar surface:
\begin{linenomath}  \begin{align}
&T(x_0, y_0, z_0) = T_{pole} \times\nonumber\\
&\left((1 - f)^{2} \sqrt{\frac{- {y_0'}^2 (1 - f)^{2} + \left({y_0'}^2 - 1\right) \left(- \omega^{2} \left({y_0'}^2 (1 - f)^{2} - {y_0'}^2 + 1\right)^{\frac{3}{2}} + 1\right)^{2}}{\left(- {y_0'}^2 (1 - f)^{2} + {y_0'}^2 - 1\right)^{3}}}\right)^{\beta}
\quad,
\end{align}  \end{linenomath}
where we define the re-scaled vertical coordinate
\begin{linenomath}  \begin{align}
    y_0' \equiv \frac{y_0}{1-f}
    \quad.
\end{align}  \end{linenomath}
%
%
%
Finally, we may use Planck's law to obtain a closed-form expression for the specific intensity at any point on the surface of the star:
\begin{linenomath}  \begin{align}
    \label{eq:I}
    I_\lambda(x_0, y_0, z_0) = \frac{2hc^2}{\lambda^5} \frac{1}{\exp\left(\frac{hc}{\lambda k_B T(x_0, y_0, z_0)}\right) - 1}
    \quad,
\end{align}  \end{linenomath}
where $h$ is Planck's constant, $c$ is the speed of light, $k_B$ is the Boltzmann constant, and $\lambda$ is the wavelength.

Computing the stellar flux during a planetary transit involves integrating Equation~(\ref{eq:I}) over the visible portion of the stellar surface projected onto the plane of the sky. While closed-form solutions exist for the case where the projected stellar surface is a disk and $I_\lambda(x_0, y_0, z_0)$ is a polynomial limb-darkening law \citep{mandel2002,Agol2020}, the present problem does not admit a simple, closed-form solution. However, since Equation~(\ref{eq:I}) is monotonic and varies quite smoothly over the surface of the star, it can be well approximated by a low-degree polynomial expansion in $y_0'$. And since the stellar surface in the transformed coordinates $(x_0, y_0', z_0)$ is the unit sphere (see Figure~\ref{fig:coordinates}), we can equivalently represent $I_\lambda$ as a low-degree expansion in the spherical harmonic basis up to degree $l_\mathrm{max}$:
\begin{linenomath} \begin{equation}
    I_\lambda(x_0, y_0, z_0) = \bvec{\tilde{y}}^\top(x_0, y_0', z_0)\,\bvec{y},
\end{equation} \end{linenomath} 
where $\bvec{y}$ is a vector of $(l_\mathrm{max} + 1)^2$ spherical harmonic coefficients representing the stellar surface intensity and $\bvec{\tilde{y}}$ is the spherical harmonic basis evaluated at a particular point $(x_0, y_0', z_0)$ on the unit sphere \citep{starry2019}. To find $\bvec{y}$, we evaluate $I_\lambda$ at a set of points spanning the stellar surface, such that we may write
\begin{linenomath} \begin{equation}
    \bvec{i}_\lambda = \bvec{\tilde{Y}}\,\bvec{y},
\end{equation} \end{linenomath} 
where $\bvec{i}_\lambda$ is the vector of specific intensities at each of our sampling points and $\bvec{\tilde{Y}}$ is the matrix whose rows are the spherical harmonic basis vectors $\bvec{\tilde{y}}^\top$ evaluated at each of those points. The spherical harmonic representation of our surface can then be computed as
\begin{linenomath} \begin{equation}
    \bvec{y} = \bvec{\tilde{Y}}^+ \bvec{i}_\lambda
    \quad,
\end{equation} \end{linenomath} 
where
\begin{linenomath} \begin{equation}
    \bvec{\tilde{Y}}^+ = (\bvec{\tilde{Y}}^\top\bvec{\tilde{Y}} + \epsilon \mathbf{I})^{-1} \bvec{\tilde{Y}}^\top
\end{equation} \end{linenomath} 
is the pseudoinverse of $\bvec{\tilde{Y}}$,
$\epsilon$ is a small constant (typically of order ${\sim}10^{-9}$) included for numerical stability, and $\bvec{I}$ is the identity matrix.%
\footnote{In practice, since the surface intensity is independent of $x_0$ and $z_0$, it is convenient to perform the spherical harmonic expansion in the polar frame, in which only the radial ($m = 0$) modes are nonzero. This is equivalent to expanding Equation~(\ref{eq:I}) in terms of one-dimensional Legendre polynomials in $y_0'$. To satisfy the Nyquist criterion, we require only $2l_\mathrm{max}$ sampling points in $y_0'$, where $l_\mathrm{max}$ is the degree of the expansion; this procedure is therefore extremely fast.}
Typically, we find that values of $l_\mathrm{max}$ between 2 and 6 are sufficient to approximate the gravity darkening profile of the star for most applications; see \S\ref{sec:performance} for details.

\section{Integrating over an Oblate Star} 
\label{sec:integration}

Taking into account the oblateness of rapidly-rotating stars in a transit model requires discarding the assumption that stars are circles in projection. Instead, we model a rapidly-rotating star as an oblate spheroid, which results in an ellipse in projection. Until now, this fact has precluded the use of analytic transit models such as those in \citet{mandel2002} for oblate stars. As we mentioned previously, the oblate transit model is typically obtained via computationally expensive 2D numerical integration methods, which can lead to transit fits that take up to several days to run \citep{ahlers2014}. 
In this section, armed with our spherical harmonic expansion of the intensity profile, $\bvec{y}$, and the integration algorithms presented in \citet{starry2019}, we derive a new semi-analytic integration scheme that greatly improves upon existing methods in both efficiency and computational precision.

\begin{figure*}[t!]
\centering
    \includegraphics[width=0.75\textwidth]{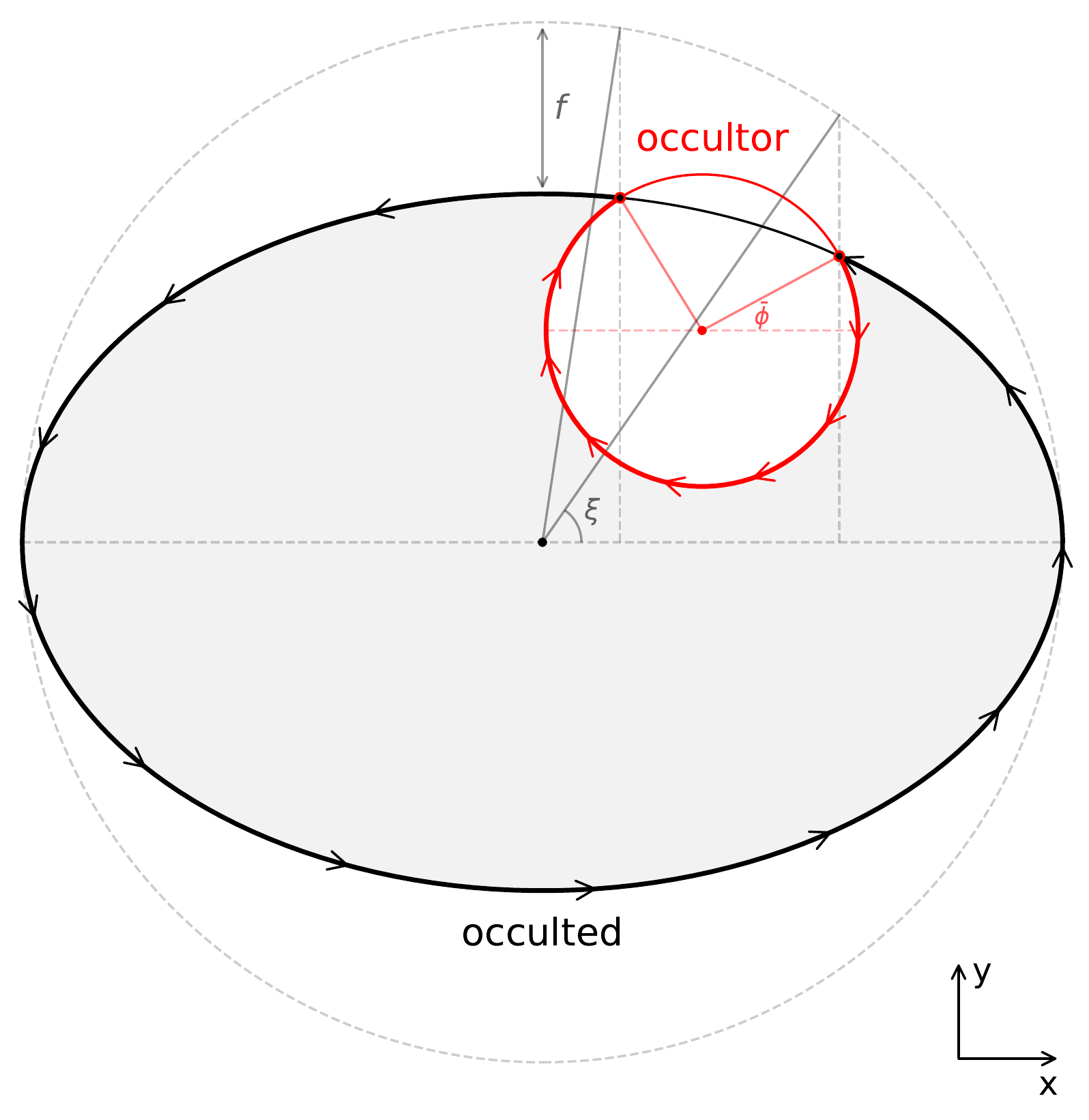}
\caption{Geometry of the occultation problem. The star is shown as a projected ellipse in black; the planet as a circle in red. The bold regions represent the boundary of integration for the planet and star respectively---together, they form a closed region around which the line integral is performed in the direction indicated by the arrows. The grey line from the center of the star to the outer bounding circle represents the angular parameter $\xi$, which is used to specify the integration limits for the integral along the stellar limb. The red line from the center of the planet to the outer planet radius represents the angle $\bar{\phi}$, which specifies the limits of integration for the integral along the planet limb.}
\label{fig:intbounds}
\end{figure*}

We start by calculating the parameters of the occulted body in projection. While the flattening of the ellipsoid is $f$ (Equation~\ref{eq:fdef}), the flattening of the ellipse \emph{in projection} can vary between $f$ (if the star is viewed at an inclination $i_*=90^\circ$) and zero (if the star is viewed at an inclination $i_*=0^\circ$); see Figure~\ref{fig:rrstar}.
As in \citet{barnes2009}, we find that the semi-minor axis of the stellar ellipse in projection is given by $1 - f'$, where the semi-major axis is unity and the effective flattening of the projected ellipse is
\begin{linenomath} \begin{equation}
    f' = 1 - \sqrt{(1-f)^2\cos^2 i_* + \sin^2 i_*}
    \quad,
\end{equation} \end{linenomath} 
where $i_*$ is the inclination of the stellar spin axis away from the line-of-sight.
We then orient our coordinates such that the semi-major axis of the projected stellar ellipse is aligned with the $x$-axis and the semi-minor axis is aligned with the $y$-axis; see Figure~\ref{fig:intbounds}.
In this frame, the occultor (planet) of radius $r_\mathrm{o}$ is at coordinates $(x_\mathrm{o}, y_\mathrm{o})$ on the plane of the sky.

We compute the total (normalized) flux $\mathcal{F}$ measured by the observer in the same way as in the \starry package \citep{starry2019}:
\begin{linenomath}  \begin{align}
    \mathcal{F} = \bvec{s^{\boldsymbol{\top}}} \bvec{A} 
    \bvec{R'}
    \bvec{R} \bvec{y}
    \quad,
\end{align}  \end{linenomath}
where, from right to left, $\bvec{y}$ is the spherical harmonic representation of the stellar intensity profile in some fixed reference frame (\S\ref{sec:gravdark}),
$\mathbf{R}$ and $\mathbf{R'}$ are spherical harmonic rotation matrices that orient the star at the correct inclination, obliquity, and angular phase at the time of the observation,
$\bvec{A}$ is a change-of-basis matrix from spherical harmonics to a convenient polynomial basis in which the integrals are computed, and
$\bvec{s} = \bvec{s}(x_\mathrm{o}, y_\mathrm{o}, r_\mathrm{o}, f')$
is the value of the two-dimensional integral of each of the terms in that basis over the unocculted portion of the projected stellar ellipse.

We devote the Appendix to the computation of the vector $\bvec{s}$ for the case of an oblate star; here we briefly summarize our method. As in the spherical case \citep{starry2019}, it entails first finding the points of intersection between the limb of the occultor and the limb of the star, which in the present case corresponds to the points of intersection of a circle and an ellipse. \citet{starry2021} showed that this corresponds to a quartic equation, whose solution can be written in closed form but in practice is faster (and more numerically accurate) to solve numerically. Next, the two-dimensional integral over the unocculted portion of the projected surface of the star is transformed into a one-dimensional line integral along the boundary of that region using Green's theorem. In general, this boundary consists of two paths, one around the limb of the star (thick black curve in Figure~\ref{fig:intbounds}) and one around the limb of the planet (thick red curve). The limits of integration for both integrals are the angular coordinates $\xi$ and $\bar{\phi}$ of the points of intersection measured from the center of the star and the planet, respectively (see the Figure).
As we show in the Appendix, most of the integrals needed to compute $\bvec{s}$ may be solved in closed form; however, we were unable to find a closed form solution to the integral along the planet limb for approximately half of the terms in the integration basis. We therefore solve these integrals using an efficient implementation of Gaussian quadrature.

Finally, we also allow for limb darkening by expressing the quadratic limb darkening law as an (exact) $l_\mathrm{max} = 2$ spherical harmonic expansion, which we fold into the matrix $\bvec{A}$ as a multiplicative function on the specific intensity profile of the star; see \citet{starry2019} and \citet{starry2021} for details.

The result is a fast and accurate algorithm for computing the flux of a spherical planet occulting an oblate, gravity- and limb-darkened star whose major and minor axes are aligned at arbitrary angles with respect to the planet (and the observer). We implement this algorithm in the \starry package along with expressions for the derivatives of the model with respect to all of the inputs for plug-and-play use with gradient-based inference schemes such as gradient descent optimization or Hamiltonian Monte Carlo (HMC).
In the following section we discuss the performance of the model in terms of accuracy, precision, and computational speed.

\section{Performance of the model} 
\label{sec:performance}

We tested the accuracy and speed of the \starry implementation of our transit model model against existing integration methods. To do this, we constructed a light curve model based on 2D numerical integration. We created a surface map following the method presented in \S\ref{sec:gravdark}. We then discretized the stellar surface under the planet into a grid of pixels at varying resolutions, where we define the resolution of the grid as the number of pixels on a side. Summing the intensities in each of these pixels yields a quantity proportional to the occulted flux, which is then subtracted from the total stellar flux to obtain the value of the transit model at any particular time.

\begin{figure}[t!]
\begin{centering}
\includegraphics[width=1.0\textwidth]{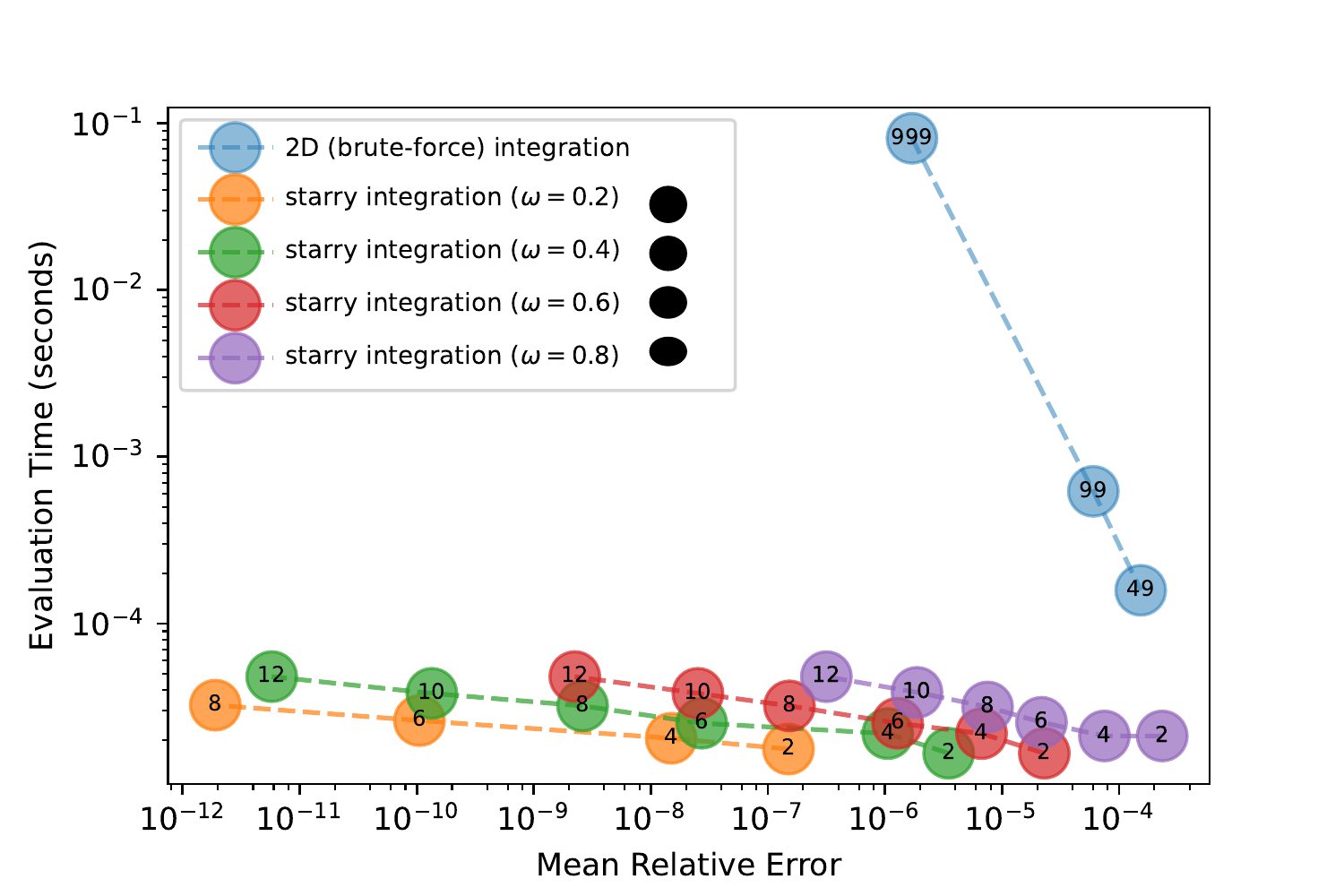}
\caption{Evaluation time in seconds versus mean relative error for both the \starry algorithm and 2D numerical integration. Each dot represents one set of hyperparameters; in the case of 2D integration the resolution of the grid and in the \starry case $l_\mathrm{max}$, the degree of the spherical harmonic expansion used to represent the gravity darkening function. The relevant value is included as a number within each dot (i.e., for 2D integration we show grid resolutions of 49, 99, and 999). Dashed lines connect dots which are similar except with different resolutions/$l_\mathrm{max}$. For the \starry integration, we also show different values of the rotation rate $\omega$ since the error of a given spherical harmonic representation scales with $\omega$. For reference, we show the oblateness of a star with the corresponding value of $\omega$ in the legend. We only show one line for 2D integration as the error does not significantly differ with $\omega$. The \starry integration is faster for all useable resolutions with 2D integration and in most cases several orders of magnitude more precise.}
\label{fig:speedtest}
\end{centering}
\end{figure}

We then tested each implementation for both runtime and accuracy. The flux for a variety of grid resolutions (for 2D numerical integration) was first computed 20 times on a standard MacBook Pro laptop with an i7 processor and the average runtime was measured. Then, the same was done for 100 iterations of the \starry solver at each of several values of the spherical harmonic degree $l_\mathrm{max}$ of the gravity darkening profile expansion and the dimensionless stellar angular velocity $\omega$ (see Figure~\ref{fig:speedtest}). 

To compute the error associated with each integration scheme, we first note the dominant sources of error in each. In the case of the 2D integration, the dominant source of error is from the discretization itself. However, in the \starry implementation, we represent gravity darkening using a finite spherical harmonic expansion, and hence the dominant source of error is expected to be with the accuracy of the spherical harmonic expansion. At low rotation rates, the star can be represented very accurately with a spherical harmonic expansion of low degree. However, as the rotation rate of the star increases, higher degree expansions are needed to adequately fit the steeper curvature of the gravity darkening function. We therefore test the error for a variety of rotation rates $\omega$ and display them all in Figure~\ref{fig:speedtest}. 

To evaluate the error, we randomly drew 150 planet positions and radii and evaluated the flux with both the 2D integration and the semi-analytic algorithm derived in this work. We then found the difference from the flux at a very high $l_\mathrm{max}$ ($l_\mathrm{max}=20$) to quantify the error. This was repeated for 4 different values of $\omega$, 6 different values of $l_\mathrm{max}$ and 3 different grid resolutions for the 2D integration and the average error for each was recorded.

We note that while in Figure~\ref{fig:speedtest} we are measuring the error against our own implementation at a high $l_\mathrm{max}$, the goal of this was to evaluate the accuracy of our method compared to existing techniques. Furthermore, we note that our values match the 2-D brute-force integration to within the discretization error of the latter, and as shown in Figure~\ref{fig:speedtest} the error of the numerical solution slowly approaches zero as we increase the grid resolution, meaning the two models do asymptotically agree. We could have computed the error for higher grid resolutions but were limited by the large amount of memory required for a high resolution grid and the length of time it would take to generate enough such models to evaluate their error. 

\begin{figure*}[ht!]
\centering
    \includegraphics[width=1.1\textwidth]{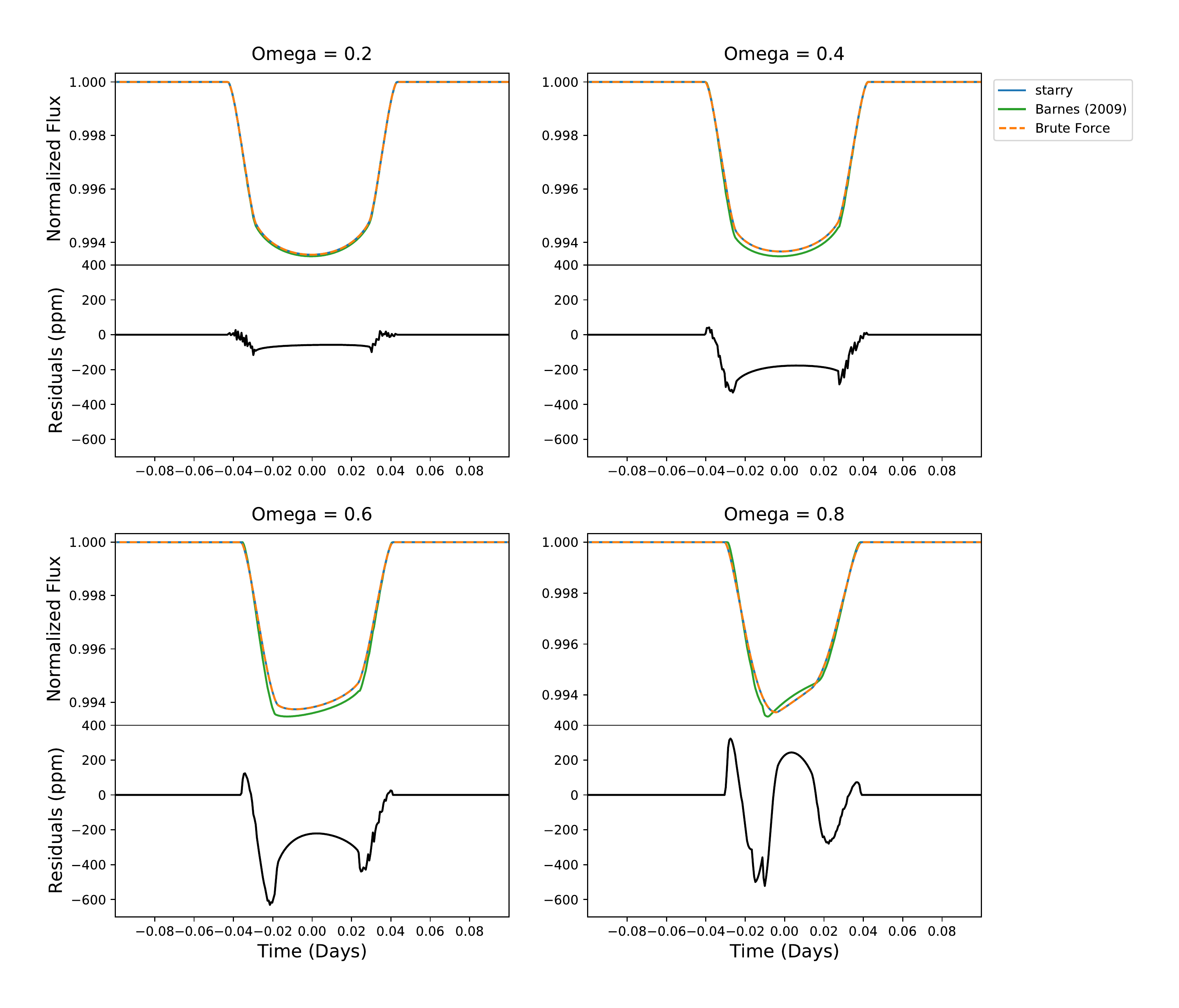}
\caption{Comparisons of light curves at different $\omega$ generated by \starry at $l_\mathrm{max}$=14, \citet{barnes2009} as implemented in \textsf{PyTransit}, and a brute force algorithm. The residuals show the difference between the \starry light curve and \textsf{PyTransit} generated light curves.}
\label{fig:starryvsbarnes}
\end{figure*}

Additionally, in order to benchmark the accuracy of \starry against other implementations of gravity-darkened and oblate transit models in the literature, we compared light curves generated by \starry, the implementation of \citet{barnes2009} in \textsf{PyTransit}, and a custom brute force integration. We generated synthetic light curves for values of $\omega$ between 0 and 1, with the resolution of the stellar grid set to N = 500 for \textsf{PyTransit} and the brute force integration and $l_\mathrm{max}$=14 for \starry. We note that all three methods visually match at low values of $\omega$. The error in this near-spherical regime is largest during ingress and egress,  indicating that the residuals are likely dominated by error introduced by the discretization in \textsf{PyTransit}. Indeed, the amplitude of the error (25ppm) matches to order of magnitude the expected error given the resolution of the \textsf{PyTransit} model (see Figure~\ref{fig:speedtest}). However, at higher $\omega$, while our custom brute force method and \starry match, there are much larger errors in the shape and depth of the \textsf{PyTransit} light curve that are unexplained by discretization.

Given that the gravity darkening effect is usually at the $\sim 10^{-4}$ level for most currently known rapidly rotating planet hosts \citep{masuda2015}, we chose resolutions for the 2D integration that would produce errors smaller than this level. We find that for $\omega < 0.2$ an $l_\mathrm{max}$ of 2 is sufficient to reach $\sim 10^{-6}$ accuracy. For $\omega = 0.8$, however, an $l_\mathrm{max} \geq 10$ is needed to reach the same order of magnitude of error. Given that the fastest known star with a substellar companion has a $v \sin i_{*}$ of $193.7$ km/s \citep{dholakia2019}, which corresponds to $\omega = 0.52$ for an equator-on star, $l_\mathrm{max} \leq 6$ will usually be sufficient for the precision of current instruments.  

Our model implemented in the \starry package is at least one order of magnitude faster and orders of magnitude more precise than common grid resolutions used to model gravity darkening and oblateness in exoplanet transits in the literature (i.e 125x125 in \citealt{lendl2020}). The model shows a good scaling with $l_\mathrm{max}$, allowing steep gravity darkening for stars with $\omega$ greater than 0.5 to be modeled straightforwardly. Lastly, the semi-analytic model implemented in \starry computes gradients using autodifferentiation, allowing derivatives to be computed for gradient-based descent and other optimization routines. This makes the model well suited for posterior inference. 

\section{Wasp 33's gravity-darkened and oblate model} \label{sec:wasp33}

We also tested our implementation of the algorithm derived in this paper on the TESS light curve of WASP-33, which has a host displaying rapid rotation of $v\sin i_* = 86.5^{+0.37}_{-0.32}$ km/s \citep{johnson2015}, and is hence expected to be significantly oblate and gravity-darkened. As found by \citet{herrero2011}, the host star is a $\delta$-Scuti with pulsations that can distort the transit signal. \citet{vonessen2020} perform a subtraction of pulsations in the light curve of WASP-33 and fit the transit to a standard, non-gravity-darkened model. They note an asymmetry in transit which they attribute to improper treatment of pulsations. Given the host star's rapid rotation, we argue that this asymmetry is due to the gravity darkening effect instead, and use this to constrain the system's true spin-orbit angle, among other system parameters. 

We started with the 2-minute cadence PDCSAP light curve of WASP-33 from TESS Sector 18. We then used the package \textsf{lightkurve} \citep{lightkurve} to perform an initial detrending.  As such, we took care to subtract the pulsation signal to recover the transit light curve with any features of gravity darkening and oblateness. We found that the subtraction method in \citet{vonessen2020} was effective in removing the pulsation signal but also reduced the asymmetry in transit due to the subtraction of a standard transit model from the TESS light curve before detrending, which allowed the asymmetry to be removed by the detrending process. 

We undertook a more conservative approach to the detrending that was agnostic to the signal in transit. We first masked out the data in transit. Then, following the LASR method of \citet{ahlers2018}, we constructed a model for the pulsations by minimizing the height of each peak in a periodogram widow centered around the peak. We started with a periodogram in the range 5-50 days$^{-1}$ and constructed windows around each peak with a power greater than $2\times 10^{-4}$. Each window was centered in a region of 0.08643 days$^{-1}$ around the peak and contained 100 sample points. Using the peak height as a goodness-of-fit metric, we then fit every peak in this window jointly with a sum of sines model using L-BFGS-B in \textsf{scipy} for the gradient descent. We iterated this process 3 times until the all remaining peaks were below the threshold $10^{-4}$. In total, 11 pulsation modes were found.

\begin{figure*}[t!]
\begin{centering}
    \includegraphics[width=0.85\textwidth]{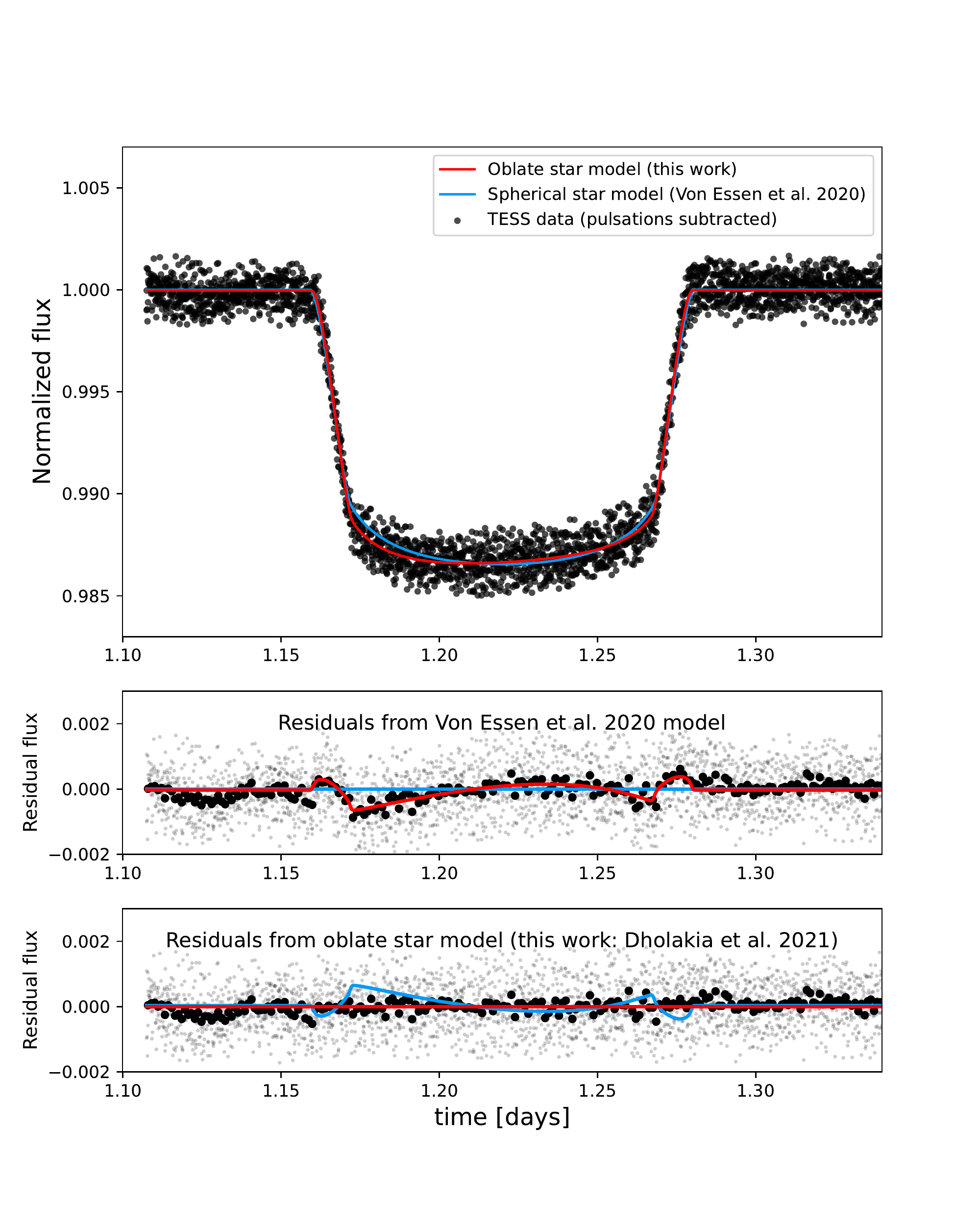}
\caption{Phase-folded TESS light curve of WASP-33 (left) with pulsations removed. The black line is the best fitting standard transit model from \citet{vonessen2020} and the red line is the best fitting oblate, gravity-darkened model from an MCMC fit. The binned residual for both models are shown, with each model overplotted on the residuals. Unbinned residuals are shown in the background in light grey.}
\label{fig:wasp33}
\end{centering}
\end{figure*}

We then extended this sum-of-sines model to the data in transit. The detrended lightcurve shows a subtle but visible slope towards the end of transit (see Figure~\ref{fig:wasp33}).

We then used our \starry model for oblate stars with gravity-darkened transits along with the package pymc3 \citep{pymc3} to derive posterior constraints on the parameters of interest via No U-Turn Sampling (NUTS) \citep{hoffman2011}. We adopted the priors for the stellar mass, radius and limb-darkening parameters from \citet{vonessen2020} and fixed the gravity darkening coefficient $\beta$ and at 0.23 (see \citet{ahlers2020a} for discussion on the risks of varying $\beta$). We used uniform priors for the impact parameter and projected spin-orbit angle in the range $-1<b<0$ and $-180^{\circ}<\lambda<180^{\circ}$, respectively. We assumed an isotropic prior
\begin{linenomath}  \begin{align}
    p(i_*)\mathrm{d}i_* = \sin i_* \mathrm{d}i_*
\end{align}  \end{linenomath}
for the stellar inclination $i_*$ and allowed the dimensionless angular velocity of the star $\omega$ (see Equation~\ref{eq:omega}) to vary with stellar inclination to match the observed $v\sin i_* = 86.5^{+0.37}_{-0.32}$ km/s \citep{johnson2015}. We note that for stellar inclinations closer to 0 (pole-on), $\omega$ becomes higher for a given $v\sin i_*$, making the gravity darkening more noticeable.

To generate model light curves that match the TESS bandpass, we used the ability of \starry to generate multiwavelength maps. We binned the TESS bandpass function \citep{ricker2014} into 30 wavelength bins and calculated a surface map for WASP-33 at each of these wavelengths by plugging the wavelength into the Planck law as described in \S\ref{sec:gravdark}. Then, we multiplied the surface map intensity by the height of the TESS bandpass at that wavelength. Each flux value was computed by integrating across the 30 wavelengths bins. We note that the integrals in \S\ref{sec:integration} and the Appendix must only be performed once across all wavelengths; as such this process is very scaleable in \starry as we increase the number of wavelengths. We also oversampled the model light curve by 5x with an exposure time of 2 minutes to avoid integration-time artifacts from affecting our posterior samples. We ran 20 chains with 1000 steps and discarded the first 500 as tuning. 

\begin{figure}[t!]
\gridline{\fig{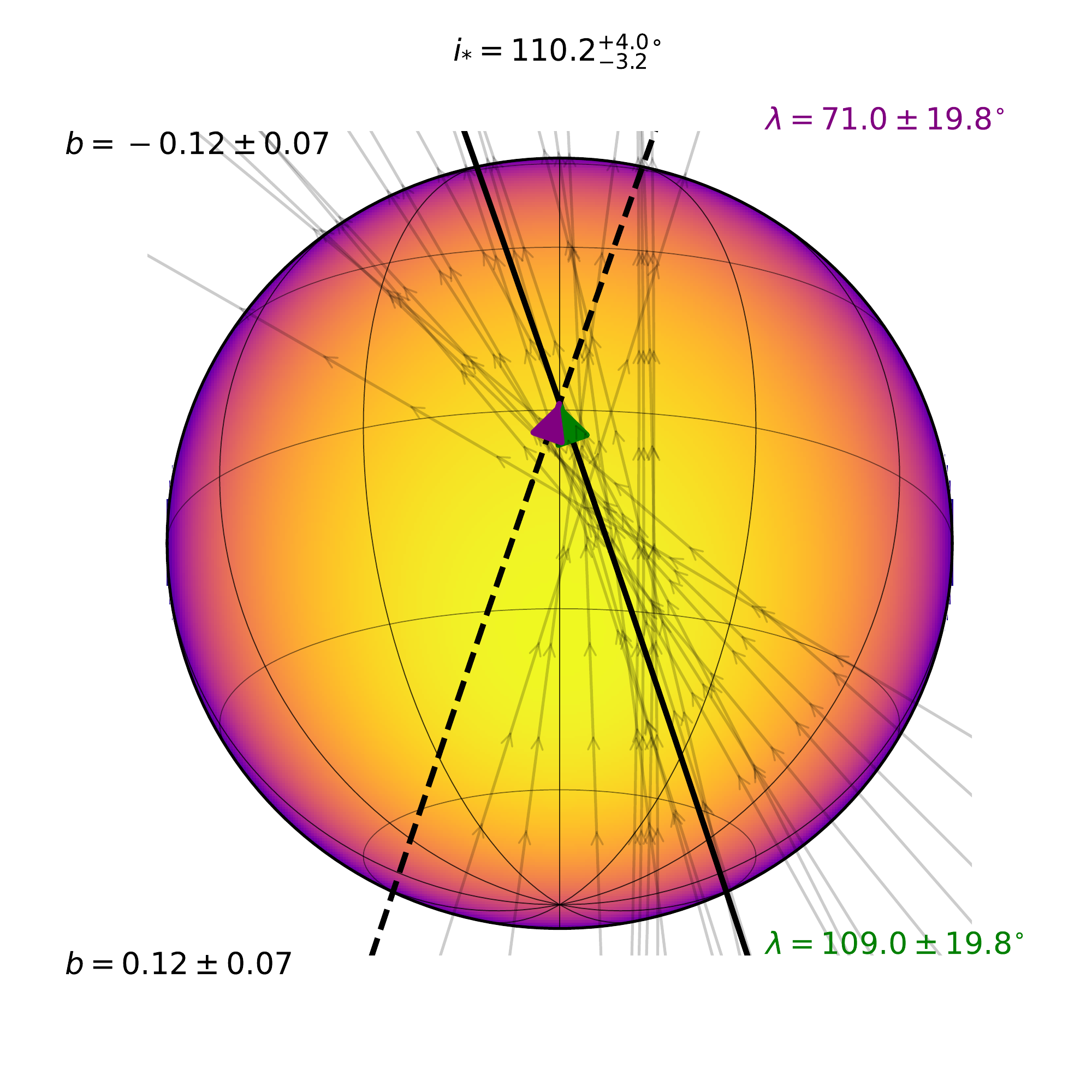}{0.5\textwidth}{}
          \fig{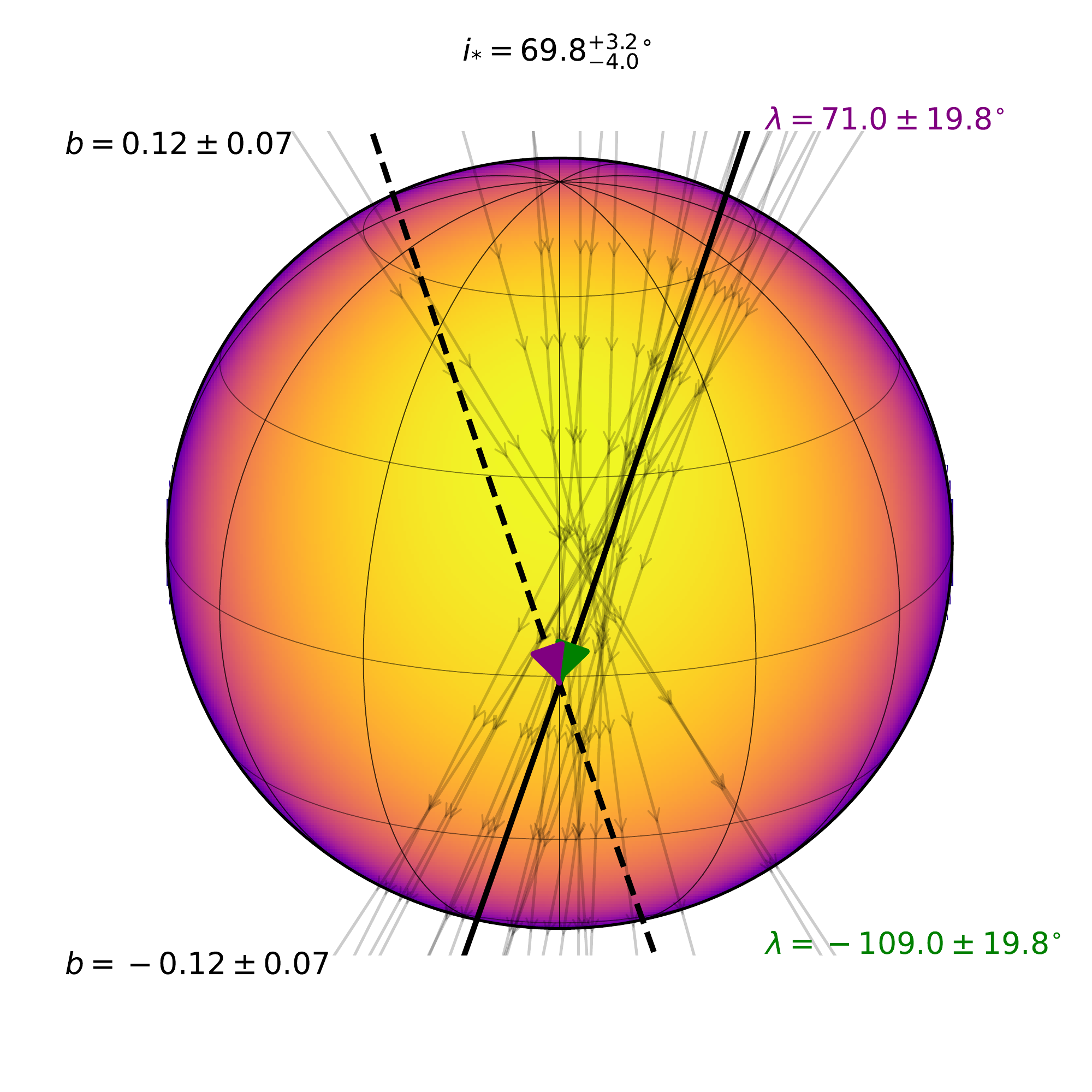}{0.5\textwidth}{}}
\caption{Visualizing the possible paths of WASP-33b crossing the stellar disk of its host star, showing the TESS light curve constraint on the system geometry. The stellar spin axis points up in both figures and the star is shown as seen by the TESS bandpass with the median value of limb darkening, polar temperature, $\omega$, and $i_{*}$ from the fit. There are two stellar inclinations which can produce the observed light curve, one with the north pole of the star visible (left) and the other with the south pole (right). The best fitting transit path is shown in black, with an alternate transit path shown as a dashed line. 20 randomly drawn posterior samples for the transit path are shown in each figure. A given value of ($b, \lambda, i_{*}$) is indistinguishable from the corresponding ($-b, -\lambda, -i_{*}$), both with transit photometry and Doppler tomography; each of these pairs of transit paths are shown with arrows of the same color across the left and right figures. Doppler tomography of WASP-33, which shows a retrograde orbit ($|\lambda|>90^{\circ}$), precludes the purple paths in both left and right figures, which are prograde. Using the expected progression of the impact parameter from nodal precession allows us to adopt the green path from the left figure. Figure layout is inspired from Figure~6 in Hooton et al. in review.}
\label{fig:orbit}
\end{figure}

As we demonstrate visually in Figure~\ref{fig:orbit}, there are multiple parameters that can produce the same light curve. Rather than sampling a multimodal posterior distribution, we opt to over-constrain the fit and then use symmetry to extend the transit parameters to all allowable transit geometries. By restricting the impact parameter to be negative in the fit, we ensure that only one transit path out of the four allowed solutions is explored (see green, purple paths in Figure~\ref{fig:orbit}). We then use the finding from \citet{johnson2015} that the system is retrograde to exclude the paths where $|\lambda|=71.0^{\circ}$. There are two remaining solutions, one where ($b>0$, $\lambda>0$, $i_{*}>90^{\circ}$) and one with ($b<0$, $\lambda<0$, $i_{*}<90^{\circ}$).

As observed by \citet{johnson2015,iorio2016, watanabe2020}, WASP-33b is undergoing nodal precession due to the oblate host star. From the latest set of data from \citet{watanabe2020}, the impact parameter is reducing and can be projected to cross 0 sometime in May 2019, just prior to the TESS observations in Nov 2019. As such, we can take the negative value for b to maintain continuity with the parameterization in the Doppler tomography observations. This gives a projected spin-orbit obliquity $\lambda$ of ${-109.0^{\circ}}^{+20.2}_{-17.6}$, which is consistent with the most recent value from \citet{watanabe2020} of $-112.91\pm0.24^{\circ}$. We find a true spin-orbit angle of ${108.3^{\circ}}^{+19.0}_{-15.4}$. We note our strong constraint on the stellar inclination of $i_{*}=69.8^{+4.0}_{-3.2}$ is discrepant from \cite{watanabe2020}'s result of ${96^{\circ}}^{+10}_{-14}$ and suggest that short term precession may be affecting their dynamic constraints on the stellar inclination. We report our full constraints from the fit to the TESS data in Table~\ref{tab:data}.

\begin{deluxetable*}{lll}
\tablecolumns{3}
\tablecaption{Fit parameters for WASP-33 system with gravity darkening and oblateness.}
\tablehead{
\colhead{Parameter} & \colhead{Symbol} & \colhead{Value}
}
\startdata
limb darkening coefficient & $u_1$ & $0.209\pm0.02$ \\
limb darkening coefficient & $u_2$ & $0.217\pm0.02$ \\
planet inclination & $i_{p}$ (deg) & $91.20^{+0.8}_{-0.9}$ \\
stellar rotation rate & $\omega$ &  $0.209\pm0.006$ \\
stellar oblateness & $f$ & $0.021\pm{0.001}$ \\
stellar inclination & $i_{*}$ (deg) & $69.8^{+4.0}_{-3.2}$ \\
impact parameter & $b$ &  $-0.12^{+0.08}_{-0.08}$ \\
projected spin-orbit angle & $\lambda$ (deg) &  $-109.0^{+17.6}_{-20.2}$\\
true spin-orbit angle & $\varphi$ (deg)& $108.3^{+19.0}_{-15.4}$\\
planet-star radius ratio & $r_{\mathrm{o}}$ & $0.1088\pm0.0003$\\
stellar equatorial radius & $R_{eq}$ ($R_{\odot}$) & $1.561^{+0.008}_{-0.008}$\\
stellar polar temperature & $T_{\mathrm{pole}}$ (K)  & $7340\pm99$ \\
\label{tab:data}
\enddata
\end{deluxetable*}

\begin{figure*}[ht]
\begin{centering}
    \includegraphics[width=0.9\textwidth]{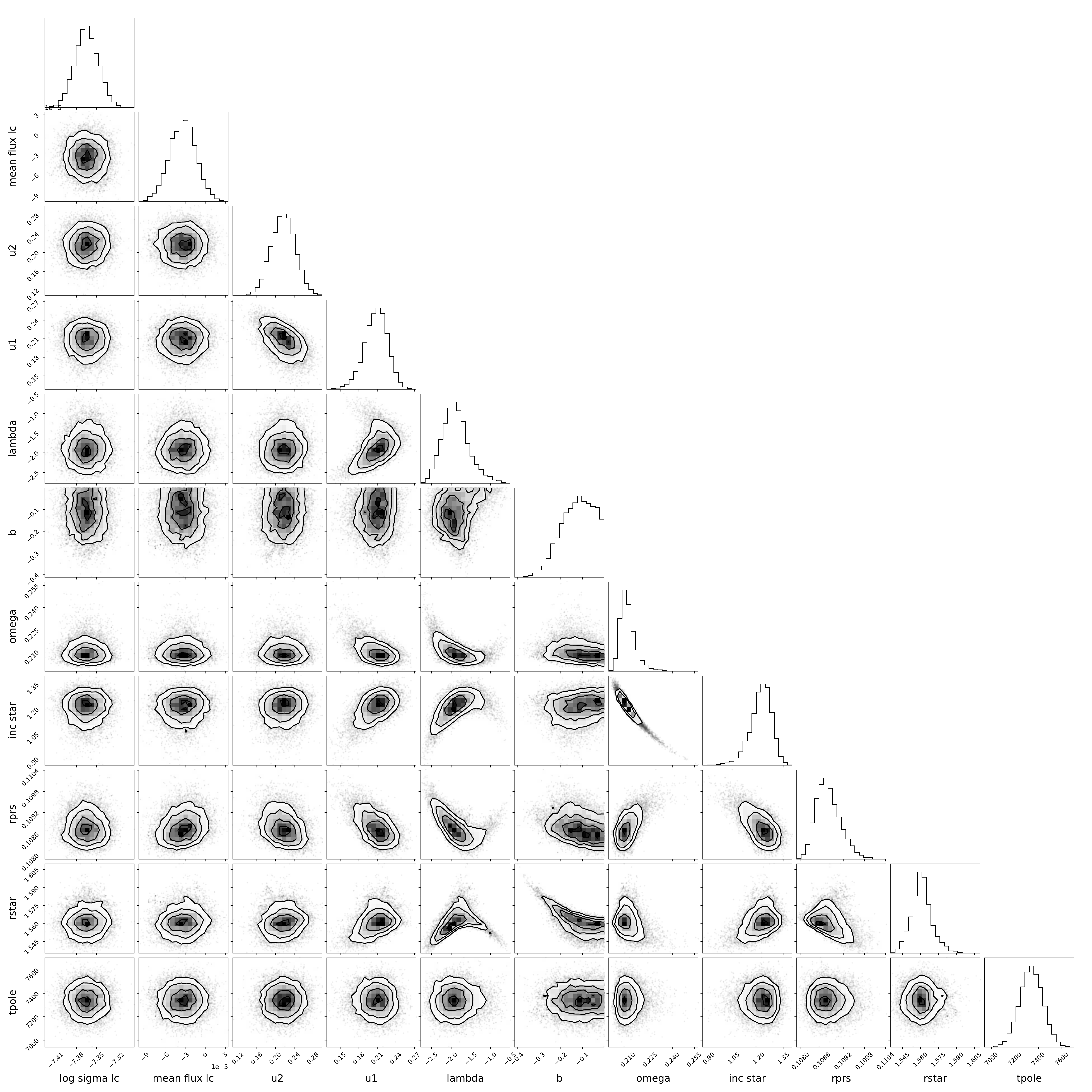}
\caption{Corner plot showing the parameter samples in the fit described in \S\ref{sec:wasp33}.}
\label{fig:corner}
\end{centering}
\end{figure*}

\section{Conclusions}  \label{sec:conclusion}

In this paper, we present a new method for the generation of efficient semi-analytic model transit light curves for rapidly rotating stars that exhibit oblateness and gravity darkening. We test the performance of the method, demonstrating several orders of magnitude better speed and precision than brute force numerical integration. Then, we apply the method to the TESS transit observation of the rapidly rotating star WASP-33 and present a novel constraint on the true spin-orbit angle of the system using photometry alone. 

We note the importance of incorporating gravity darkening and oblateness to model stars that are known to be rapidly-rotating, such as WASP-33. Asymmetry in transit is usually cited as a reason to employ models incorporating rapid stellar rotation effects. However, as shown by \citet{barnes2009}, certain configurations of the planet and star do not produce asymmetric light curves but can still contribute to a constraint on the spin-orbit angle.

Furthermore, another effect of oblateness is often overlooked as a reason for employing an oblate, gravity-darkened fit: differences in the ingress and egress duration and shape. This is caused by the planet transiting limbs of the star with different curvature during ingress and egress. We show in Figure~\ref{fig:starryvsbarnes} that current numerical models have large errors especially at ingress and egress. The precision of the semi-analytic \starry model for oblate stars allows for its use to constrain system parameters using just the shape of ingress and egress.

While we expect the effects of this to be small in case of WASP-33 due to the planet's nearly polar orbit, we can see in Figure~\ref{fig:wasp33} that the model in \citet{vonessen2020} had large deviations at ingress and egress, perhaps due to systematic errors from the spherical model fit. Our value of the planet-star radius ratio ($r_{\mathrm{o}}$ in Table~\ref{tab:data}) is correspondingly $4.3\sigma$ discrepant from the value in \citet{vonessen2020} with the same original TESS light curve. This also underscores the importance of incorporating rapid stellar rotation effects to avoid systematic errors on high precision light curves.

\subsection{Importance of posterior inference in the gravity darkening problem}
We also emphasize the use of Bayesian inference in the problem of fitting gravity-darkened and oblate transit models to data. Previously, due to the computational complexity of existing routines, many authors have used least squares optimization methods. With Bayesian inference, is also more straightforward to propagate prior available information, such as constraints from Doppler tomography or multiwavelength observations, into the posterior.

As found by \citet{johnson2014} in the case of Kepler-13Ab, the gravity darkening approach to transit fitting is prone to a host of degeneracies that can significantly affect the resulting spin-orbit angle. Parameters like $\beta$, the gravity darkening coefficient, and limb darkening parameters can greatly affect the parameters from the fit \citep{ahlers2020}. For this reason, it is often very advantegeous to use this method in tandem with any available ground based data such as Doppler tomography. In addition, recent work has suggested that multicolor photometry is also very effective at breaking these degeneracies (Hooton et al. in review). 

The existence of these degeneracies and the solution of combining multiple datasets is one important reason to use Bayesian inference as a method to perform gravity-darkened fits. Bayesian inference allows the use of Doppler tomography or other existing constraints on the spin-orbit as a prior, reducing the degeneracies from the fit. 

Another reason to use posterior inference is to expand the purview of the method to the TESS field. Every gravity-darkened and oblate transit fit to date (see \S\ref{sec:intro}) has been used on systems with visible asymmetry in the light curve. However, aligned planets do not produce visible asymmetry in the transit light curve, even when the star is rapidly rotating, leading to a bias in the systems analyzed using this method. In order to make gravity darkening feasible to perform photometric surveys of spin-orbit alignment, strong constraints should be made on alignment even when there is no asymmetry visible in the transit. This can be done using a strong constraint on the $v \sin i_*$ of the star from spectroscopy and exploring the posterior space to find the posterior probability of spin-orbit alignment.

\subsection{Future work}

The implementation of our model in \starry and its use of spherical harmonics to model the surface of celestial bodies allows for many of the assumptions made in this paper to be relaxed. For instance, we use the Von Zeipel law to model the gravity darkening effect; any other gravity darkening law such as that presented in \citet{espinosa2011} is equally straightforward to model using a spherical harmonic expansion. One can also use any relation between the rotation rate $\omega$ and the oblateness $f$. While the implementation in \starry is primarily designed to model gravity-darkened and oblate transits, the same can be applied with minor modifications to model other situations where gravity darkening and oblateness are non-negligible, such as heartbeat stars and binary systems where one star is rapidly rotating. 

TESS is projected to find ~2000 planets around A and F stars \citep{ahlers2020a}, a significant fraction of which will be rapidly rotating and amenable to constraints using the photometric method presented in this paper. Our implementation in \starry is fast and precise enough to perform posterior inference and make constraints on such planets photometrically, optionally with priors from Doppler tomography or observations from other wavelengths. Data from space based telescopes such as CHEOPS, HST and JWST may also yield precise transit observations of planets orbiting rapidly rotating stars, many of which are amenable to atmospheric characterization. With the precise and fast transit model presented in this paper and implemented in \starry, taking into account rapid stellar rotation effects with high precision transit measurements is possible. 

The software presented in this paper is open source under the MIT License and is available at \url{https://github.com/rodluger/starry}. The documentation for \starry is available at \url{https://starry.readthedocs.io}. The code used to generate the figures in the paper and derivations for some equations presented is hosted at \url{https://github.com/shashankdholakia/gravity-dark}.

\software{starry \citep{starry2019}, exoplanet \citep{exoplanet2021}, lightkurve \citep{lightkurve}, corner \citep{corner2016}, scipy \citep{scipy2020}, pymc3 \citep{pymc3}, PyTransit \citep{parviainen2015}}

\acknowledgments{We would like to thank John Ahlers, Tom Barclay, Eric Agol, and Matthew Hooton for helpful discussions while writing the manuscript. RL is supported by a Flatiron Fellowship at the Center for Computational Astrophysics, a division of the Simons Foundation.}

\newpage
\bibliography{references}

\newpage
\appendix
\label{appendix}

\section{The starry algorithm} \label{sec:greens}

In this section, we make modifications to the \starry formalism to account for stellar oblateness. It is instructive to review the algorithm for the flux from a spherical star, which we do first.

\subsection{Spherical star}
\citet{starry2019} showed that the instantaneous flux $\mathcal{F}_\sph$ observed during an occultation of a spherical star by a spherical occultor may be computed from
\begin{linenomath}  \begin{align}
    \mathcal{F}_\sph^{\,} = \bvec{s_\sph^{\boldsymbol{\top}}} \bvec{A} \bvec{R'} \bvec{R} \bvec{y}
    \quad,
\end{align}  \end{linenomath}
where, from right to left, 
$\bvec{y}$ is the vector of spherical harmonic coefficients describing the surface map of the star in some fixed reference frame,
$\bvec{R}$ is a Wigner rotation matrix that rotates $\bvec{y}$ to the sky frame given the star’s inclination, obliquity, and rotational phase \citep[Appendix C in][]{starry2019}, 
$\bvec{R'}$ is a second Wigner rotation matrix that rotates the star on the plane of the sky into a frame in which the occultor lies along the $+y$-axis \citep[Equation 23 in][]{starry2019}, and
$\bvec{A}$ \citep[Equation B13 in][]{starry2019} is the change-of-basis matrix from
the spherical harmonic basis to the \emph{Green’s basis} $\gbasis$ in which the integrals are computed \citep[Equation 11 in][]{starry2019}:
\begin{linenomath}  \begin{align}
    \label{eq:gn}
    \tilde{g}_{n}(x, y) & \equiv
    \begin{dcases}
        \frac{l-m+2}{2}x^\frac{l-m}{2} y^\frac{l+m}{2}
         & \qquad l + m \, \text{even}
        \\[1em]
        z
         & \qquad l = 1, m = 0
        \\[1em]
        3x^{l-2}yz
         & \qquad m = l - 1, \, l \, \mathrm{even}
        \\[1em]
        \bigg(
        -x^{l-3} + x^{l-1} + 4x^{l-3}y^2
        \bigg)z
         & \qquad m = l - 1, \, l \, \mathrm{odd}
        \\[1em]
        \frac{1}{2}
        \bigg(
        (l-m-3) x^\frac{l-m-5}{2} y^\frac{l+m-1}{2}
        \\
        \,\,\,\,\,\, - \
        (l-m-3) x^\frac{l-m-5}{2} y^\frac{l+m+3}{2}
        \\
        \,\,\,\,\,\, - \
        (l-m+3) x^\frac{l-m-1}{2} y^\frac{l+m-1}{2}
        \bigg)z
         & \qquad \text{otherwise}
        \quad,
    \end{dcases}
\end{align}  \end{linenomath}
where
\begin{linenomath}  \begin{align}
    \label{eq:lofn}
    l &= \floor{\sqrt{n}} 
    \\
    \label{eq:mofn}
    m &= n - \floor{\sqrt{n}}^2 - \floor{\sqrt{n}}
\end{align}  \end{linenomath}
and
\begin{linenomath}  \begin{align}
    \label{eq:z}
    \z \equiv \sqrt{1-\x^2 - y^2}
\end{align}  \end{linenomath}
and $x$ and $y$ are the right-handed Cartesian coordinates on the plane of the sky, where $\yhat$ points up, $\xhat$ points to the right, and $\zhat$ points toward the observer.
Finally, the vector $\bvec{s_\sph^\top}$ is the \emph{solution vector}, the integral of each of the
terms in the Green's basis over the projected visible portion of the spherical star:
\begin{linenomath}  \begin{align}
    \bvec{s_\sph} \equiv \iint \gbasis (\x, \y)  \, \dd S
    \quad.
\end{align}  \end{linenomath}
\citet{starry2019} use Green's theorem to cast this two-dimensional surface integral as a line integral over the boundary of the integration region. Thus the $n^\mathrm{th}$ term of the solution vector may be written as
\begin{linenomath}  \begin{align}
    \label{eq:greens}
    s_{\sph,n} &=
    \iint \gbasisn (\x, \y) \, \dd S
    \nonumber\\
    &=
    \oint \bvec{G}_{\sph,n} (\x, \y) \cdot \dd \bvec{r}
    \quad,
\end{align}  \end{linenomath}
where $\bvec{G}_{\sph,n}$ is the anti-exterior-derivative of $\tilde{g}_n$
\citep{Pal2012,starry2019} and $\dd \bvec{r}$ is the differential element along the integration path.
The vector function $\bvec{G}_{\sph,n} (\x, \y)$ is arbitrary, but must satisfy the condition that the exterior derivative of $\bvec{G}_{\sph,n} (\x, \y)$ is equal to $\gbasisn (\x, \y)$:
%
\begin{linenomath}  \begin{align}
    \frac{d (\mathbf{G}_{\sph,n} \cdot \yhat)}{dx} 
    - 
    \frac{d (\mathbf{G}_{\sph,n} \cdot \xhat)}{dy}
    =
    \gbasisn(x, y)
    \quad.
\end{align}  \end{linenomath}
For a spherical star, this is given by Equation~(34) in \citet{starry2019}. The one-dimensional integral in Equation~(\ref{eq:greens}) may then be computed in closed form (see Appendix D in \citealt{starry2019}).

\subsection{Oblate star}
\begin{figure}[t!]
\begin{centering}
\includegraphics[width=0.65\linewidth]{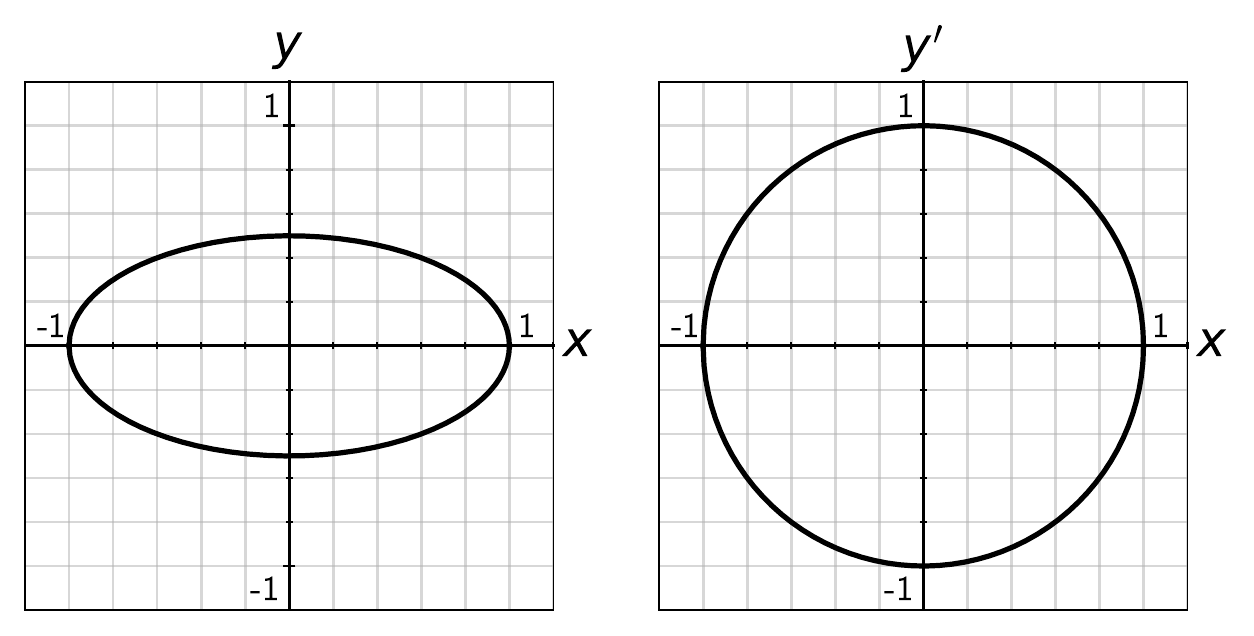}
\caption{\emph{Left}: The surface of an oblate ($f'=0.5$) spheroid projected onto the plane of the sky, defined by a right-handed Cartesian coordinate system $(x, y)$. \emph{Right}: The transformed sky coordinates $(x, y')$ used to compute the flux, in which the ellipse is stretched into a circle via the transformation $y' \equiv \nicefrac{y}{(1 - f')}$.}
\label{fig:coordinates}
\end{centering}
\end{figure}

The flux computation in the oblate case is similar to the above:
\begin{linenomath}  \begin{align}
    \label{eq:starry_obl}
    \mathcal{F}_\obl^{\,} = \bvec{s_\obl^{\boldsymbol{\top}}} \bvec{A} 
    \bvec{R'}
    \bvec{R} \bvec{y}
    \quad,
\end{align}  \end{linenomath}
In order to compute the oblate solution vector $\bvec{s_\obl^{\boldsymbol{\top}}}$,
we must make changes to account for the different metric along the surface of the star, which in projection is now an ellipse with flattening $f'$ (left panel of Figure~\ref{fig:coordinates}). Without loss of generality, let us normalize distances such that the semi-major axis of the projected ellipse is unity; let us also align it with the $x$-axis. The semi-minor axis is then equal to $1 - f'$ and is aligned with the $y$-axis.

Since the Green's basis is defined on the unit disk, we introduce the transformed coordinate system
\begin{linenomath}  \begin{align}
    \label{eq:yprime}
    x' &\equiv x
    \\
    y' &\equiv \frac{y}{1 - f'}
    \quad.
\end{align}  \end{linenomath}
In this transformed coordinate system, the stellar ellipse is stretched out into the unit circle (right panel of Figure~\ref{fig:coordinates}). The equation defining the anti-exterior-derivatives of the Green's basis is now
\begin{linenomath}  \begin{align}
    \label{eq:exterior_obl}
    \frac{d (\mathbf{G}_{\obl,n} \cdot \yhat)}{dx} 
    - 
    \frac{d (\mathbf{G}_{\obl,n} \cdot \xhat)}{dy}
    =
    \gbasisn(x, y')
    \quad,
\end{align}  \end{linenomath}
where we must also take care to replace $z$ with 
\begin{linenomath}  \begin{align}
    \label{eq:zprime}
    \z' \equiv \sqrt{1-\x^2 - y'^2}
\end{align}  \end{linenomath}
in Equation~(\ref{eq:gn}). One particular form for $\bvec{G}_{\obl,n}$ that satisfies Equation~(\ref{eq:exterior_obl}) is
\begin{linenomath}  \begin{align}
    \label{eq:Gobln}
    \bvec{G}_{\obl,n} (\x, \y) &=
    \begin{dcases}
        \x^{\frac{l - m + 2}{2}}
        y'^{\frac{l + m}{2}}
        \,\yhat
            & \qquad l + m \, \mathrm{even}
        \\[1em]
        \frac{1-z'^3}{3(1-z'^2)}(-\y \, \xhat + \x \, \yhat)
            & \qquad l = 1, \, m = 0
        \\[1em]
        (1-f') \x^{l-2}
        \z'^3
        \,\xhat
            & \qquad m = l - 1, \, l \, \mathrm{even}
        \\[1em]
        \x^{l-3}
        \y
        \z'^3
        \,\xhat
         & \qquad m = l - 1, \,
                  l \, \mathrm{odd}
        \\[1em]
        \x^{\frac{l - m - 3}{2}}
        y'^{\frac{l + m - 1}{2}}
        \z'^3
        \,\yhat
            & \qquad \mathrm{otherwise.}
    \end{dcases}
\end{align}  \end{linenomath}
In the limit $f' \rightarrow 0$, this expression reduces to that in Equation~(34) of \citet{starry2019} for a spherical star.
We then compute the solution integral in the same way as before:
\begin{linenomath}  \begin{align}
    \label{eq:greens_obl}
    s_{\obl,n} &=
    \oint \bvec{G}_{\obl,n} (\x, \y) \cdot \dd \bvec{r}
    \quad.
\end{align}  \end{linenomath}
The integration path is the boundary of the region over which we wish to integrate (by Green's theorem), so this will in general consist of a path along the limb of the (spherical) occultor and a path along the limb of the (oblate) star. Thus we may write
\begin{linenomath}  \begin{align}
    \label{eq:sn}
    s_{\obl,n} &=
        \oint
        \bvec{G}_{\obl,n} (x, y)
        \cdot
        d\bvec{r}
        \nonumber \\
        &=
        \oint
        \bvec{G}_{\obl,n} (x, y)
        \cdot
        (dx \, \xhat + dy \, \yhat)
        \nonumber \\
        &=
        \int_{\phi_1}^{\phi_2}
        \left(
            \bvec{G}_{\obl,n} \big(x\left(\phi\right), y\left(\phi\right)\big) \cdot \xhat \, \frac{dx(\phi)}{d\phi}
            +
            \bvec{G}_{\obl,n} \big(x\left(\phi\right), y\left(\phi\right)\big) \cdot \yhat \, \frac{dy(\phi)}{d\phi}
        \right) \, d\phi
        \nonumber\\
        &+
        \int_{\xi_1}^{\xi_2}
        \left(
            \bvec{G}_{\obl,n} \big(x\left(\xi\right), y\left(\xi\right)\big) \cdot \xhat \, \frac{dx(\xi)}{d\xi}
            +
            \bvec{G}_{\obl,n} \big(x\left(\xi\right), y\left(\xi\right)\big) \cdot \yhat \, \frac{dy(\xi)}{d\xi}
        \right) \, d\xi
\end{align}  \end{linenomath}
or
\begin{linenomath}  \begin{align}
    \label{eq:sn_pn_tn}
    s_{\obl,n} &\equiv p_{n} + t_{n}\quad,
\end{align}  \end{linenomath}
where $p_{n}$ is the line integral from $\phi_1$ to $\phi_2$ along the limb of the spherical occultor and $t_n$ is the line integral from $\xi_1$ to $\xi_2$ along the limb of the oblate star.
We parametrize the path along the limb of the occultor as
\begin{linenomath}  \begin{align}
    \label{eq:xofphi}
    x(\phi) &= r_\mathrm{o} \cos \phi  \cos\theta + (b + r_\mathrm{o} \sin \phi)  \sin\theta
    \\
    \label{eq:yofphi}
    y(\phi) &= -r_\mathrm{o} \cos \phi  \sin\theta + (b + r_\mathrm{o} \sin \phi)  \cos\theta
\end{align}  \end{linenomath}
and the path along the limb of the star as
\begin{linenomath}  \begin{align}
    \label{eq:xofxi}
    x(\xi) &= \cos \xi
    \\
    \label{eq:yofxi}
    y(\xi) &= (1 - f') \sin \xi
    \quad,
\end{align}  \end{linenomath}
where
\begin{linenomath}  \begin{align}
    \theta = \mathrm{arctan2}(x_\mathrm{o}, y_\mathrm{o})
\end{align}  \end{linenomath}
is the angular position of the planet measured clockwise from the $+y$-axis
and
\begin{linenomath}  \begin{align}
    b = \sqrt{x_\mathrm{o}^2 + y_\mathrm{o}^2}
\end{align}  \end{linenomath}
is the impact parameter; in the equations above, $(x_\mathrm{o}, y_\mathrm{o})$ is the instantaneous Cartesian position of the occultor (the planet) on the plane of the sky, and $r_\mathrm{o}$ is its radius. Note that since we account for this rotation at this step, we set $\mathbf{R'} = \mathbf{I}$ in Equation~(\ref{eq:starry_obl}).

Figure~\ref{fig:intbounds} illustrates this problem for a particular occultation configuration, with the star indicated as the black ellipse and the occulting planet as the red circle. The angles $\xi$ and
\begin{linenomath}  \begin{align}
    \label{eq:varphi}
    \bar{\phi} \equiv \phi - \theta
\end{align}  \end{linenomath}
are indicated; both angles are measured counter-clockwise from the $x$-axis. Note that while $\phi$ and $\bar{\phi}$ are proper angles (measured along the circumference of a circle), $\xi$ is an angular parameter measured along the perimeter of an ellipse, and is thus computed in the same way as the eccentric anomaly of a Keplerian orbit: it is measured along the circumference of the unit circle at the same $x$ coordinate as the corresponding point on the ellipse. The line integral of $\bvec{G}_{\obl,n}$ along the thick black curve yields $t_n$, while the line integral of $\bvec{G}_{\obl,n}$ along the thick red curve yields $p_n$; both are taken in the direction indicated by the arrows. By Green's theorem, the sum of these integrals is the surface integral of the $n^\mathrm{th}$ term in the Green's basis over the unocculted region of the star (light gray shading). These integrals are then transformed into the spherical harmonic basis, rotated, and linearly combined with the expansion coefficients $\bvec{y}$ to yield the model for the flux seen by the observer.

\section{Integration bounds} 
\label{sec:bounds}

To find the limits of integration $\phi_1, \phi_2, \xi_1, \xi_2$, we start with the equations for the ellipse and circle that represent the occulted body and occultor respectively in projection in the integration frame (see Figure~\ref{fig:intbounds}). In this frame, the equation for the boundary of the star is
\begin{linenomath} \begin{equation}
\label{eq:star_boundary}
y^2 = (1-f')^{2} \left(1 - x^{2}\right)
\end{equation} \end{linenomath} 
and the equation for the boundary of the occultor is
\begin{linenomath} \begin{equation}
\label{eq:occultor_boundary}
(y - y_\mathrm{o})^2 = r_\mathrm{o}^{2} - (x - x_\mathrm{o})^{2}
\quad.
\end{equation} \end{linenomath} 
The points of intersection between these two curves are the limits of integration for our line integrals, which we obtain by substituting Equation~(\ref{eq:star_boundary}) into Equation~(\ref{eq:occultor_boundary}) and solving for the $x$ coordinates of the intersections.
This procedure yields a quartic polynomial of the form 
\begin{linenomath}  \begin{align}
\label{eq:quartic}
\mathrm{A}x^4 + \mathrm{B}x^3 + \mathrm{C}x^2 + \mathrm{D}x + \mathrm{E} = 0
\quad,
\end{align}  \end{linenomath}
for which A, B, C, D and E are given by
\begin{linenomath}  \begin{align}
\mathrm{A} &= f'^{2} \left(f' - 2\right)^{2}
\\
\mathrm{B} &= 4 f' x_\mathrm{o} \left(f' - 2\right)
\\
\mathrm{C} &= - 2 r_\mathrm{o}^{2} + 6 x_\mathrm{o}^{2} + 4 y_\mathrm{o}^{2} + 2 y_\mathrm{o} - 2 \left(1 - f'\right)^{4} + 2 \left(1 - f'\right)^{2} \left(r_\mathrm{o}^{2} - x_\mathrm{o}^{2} - y_\mathrm{o} + 1\right)
\\
\mathrm{D} &= - 4 x_\mathrm{o} \left(- r_\mathrm{o}^{2} + x_\mathrm{o}^{2} + 2 y_\mathrm{o}^{2} + y_\mathrm{o} + \left(1 - f'\right)^{2}\right)
\\
\mathrm{E} &= r_\mathrm{o}^{4} - 2 r_\mathrm{o}^{2} x_\mathrm{o}^{2} - 4 r_\mathrm{o}^{2} y_\mathrm{o}^{2} - 2 r_\mathrm{o}^{2} y_\mathrm{o} + x_\mathrm{o}^{4} +  4 x_\mathrm{o}^{2} y_\mathrm{o}^{2} + 2 x_\mathrm{o}^{2} y_\mathrm{o} + y_\mathrm{o}^{2} \nonumber \\ & \ \ \ + \left(1 - f'\right)^{4} + 2 \left(1 - f'\right)^{2} \left(- r_\mathrm{o}^{2} + x_\mathrm{o}^{2} + y_\mathrm{o}\right)
\end{align}  \end{linenomath}
While the roots of quartic polynomials can be found in closed form, we find that it is faster and more computationally stable to solve Equation~(\ref{eq:quartic}) numerically via eigendecomposition of its companion matrix \citep[e.g.,][]{Edelman95} and root polishing using Newton's method.

Solving this quartic provides the points of intersection (if any) that can be used to construct a closed region to perform the line integral in Equation~(\ref{eq:greens}). The $x$ and $y$ values of the intersection points are used to compute the angular parameters $\xi$ and $\phi$ as described in \citet{starry2021}. In general, there may be zero, one, two, or four points of intersection.

\subsection{No points of intersection}
This case is encountered if there is no occultation, in which case
\begin{linenomath}  \begin{align}
    \phi_1 &= \phi_2 = 0 \\
    \xi_1 &= 0 \\
    \xi_2 &= 2\pi
    \quad;
\end{align}  \end{linenomath}
if the occultation is complete (the star is fully occulted by the planet or companion), in which case
\begin{linenomath}  \begin{align}
    \phi_1 &= \phi_2 = 0 \\
    \xi_1 &= \xi_2 = 0
    \quad;
\end{align}  \end{linenomath}
or if the planet occults the star but is fully enclosed within the stellar ellipse, in which case
\begin{linenomath}  \begin{align}
    \phi_1 &= 2\pi \\
    \phi_2 &= 0 \\
    \xi_1 &= 0 \\
    \xi_2 &= 2\pi
    \quad.
\end{align}  \end{linenomath}

\subsection{One point of intersection}
This case is encountered when the disk of the occultor is perfectly tangent to the boundary of the stellar ellipse, and always reduces to one of the three cases in the previous section (no occultation, complete occultation, or fully enclosed occultation.

\subsection{Two points of intersection}
This is the case during transit ingress and egress. If $x$ is a real solution to the quartic (Equation~\ref{eq:quartic}), the corresponding angles are given by
\begin{linenomath}  \begin{align}
    \phi &= \theta \pm \mathrm{arctan2}\bigg((1 - f') \sqrt{1 - x^2} \mp y_\mathrm{o}, x - x_\mathrm{o}\bigg)
    \\
    \xi &= \mathrm{arctan2}(\pm\sqrt{1 - x^2}, x)
    \quad.
\end{align}  \end{linenomath}
Whether these correspond to $(\phi_1, \xi_1)$ or $(\phi_2, \xi_2)$ is determined by the sense of integration: the integral along the boundary of the star is taken in the counter-clockwise direction, and the integral along the boundary of the planet in the clockwise direction. In practice, it is sometimes also necessary to add or subtract factors of $2\pi$ to enforce this convention.

\subsection{Four points of intersection}
This can be the case if the projected oblateness $f' \ge 1 - \sqrt{r_{\mathrm{o}}}$. This is only possible when $f'$ is large and the size of the occultor approaches the size of the occulted body, which is unlikely for an exoplanet systems but could be necessary for some binary systems. As this case requires integrating over four distinct paths, we do not at present implement it in \starry and instead raise an error if it is encountered.

\section{Solutions to the integrals} 
\label{sec:analytic}

Below we present the solution to the integrals in Equations~(\ref{eq:sn}) and (\ref{eq:sn_pn_tn}) for the computation of the solution vector $\bvec{s_\sph^{\boldsymbol{\top}}}$. Since the expression for $\bvec{G}_{\obl,n}$ has several branches (Equation~\ref{eq:Gobln}), we consider each case separately below.

\subsection{Case 1: $l + m$ even}
This case has a closed form solution for both $p_n$ and $t_n$.
The integral along the limb of the occultor is
\begin{linenomath}  \begin{align}
    p_n &=
        \int_{\phi_1}^{\phi_2}
        \x(\phi)^{\frac{l - m + 2}{2}}
        \left(\frac{\y(\phi)}{1-f'}\right)^{\frac{l + m}{2}}
        \frac{dy(\phi)}{d\phi} d\phi
        \nonumber\\
        &=
        (1-f')^{-\frac{l + m}{2}}
        \mathcal{M}_{l, \frac{l - m + 2}{2}, \frac{l + m}{2}}
        \label{eq:pn_case1}
\end{align}  \end{linenomath}
where
\begin{linenomath}  \begin{align}
    \label{eq:Mlij}
    \mathcal{M}_{l, i, j} &\equiv
        \sum_{p=0}^{l + 1}
        \sum_{q=0}^{l + 2}
        \mathcal{S}_{l,i,p+1}
        \mathcal{C}_{l,q,j}
        \mathcal{L}_{l+2-p,l+2-q}
\\
    \label{eq:Slij}
    \mathcal{S}_{l, i, j} &\equiv
        {i \choose i + j - l - 2}
        r^{l + 3 - j}
        (b \sin \theta)^{i + j - l - 2}
\\
    \label{eq:Clij}
    \mathcal{C}_{l, i, j} &\equiv
        {j \choose i + j - l - 2}
        r^{l + 2 - i}
        (b \cos \theta)^{i + j - l - 2}
\\
    \label{eq:Lij}
    \mathcal{L}_{i,j} &\equiv
    \int_{\bar{\phi}_1}^{\bar{\phi}_2} \cos^i w \sin^j w \, d w
\end{align}  \end{linenomath}
and we again make use of $\bar{\phi} \equiv \phi - \theta$ (Equation~\ref{eq:varphi}).
The integral in the last equation is the same as in Equation~(C103) of \citet{starry2021} and can be computed recursively from the lower boundary conditions
\begin{linenomath}  \begin{align}
\label{eq:L00}
\mathcal{L}_{0,0} &= \bar{\phi}_2 - \bar{\phi}_1
\\
\label{eq:L10}
\mathcal{L}_{1,0} &= \sin\bar{\phi}_2 - \sin\bar{\phi}_1
\\
\label{eq:L01}
\mathcal{L}_{0,1} &= \cos\bar{\phi}_1 - \cos\bar{\phi}_2
\\
\label{eq:L11}
\mathcal{L}_{1,1} &= \frac{1}{2}\left(\cos^2\bar{\phi}_1 - \cos^2\bar{\phi}_2\right)
\end{align}  \end{linenomath}
and the upward recursion relations
\begin{linenomath}  \begin{align}
    \label{eq:Lijupj}
    \mathcal{L}_{i,j} &=
    \frac{1}{i + j}
    \bigg(
        \cos^{i+1}\bar{\phi}_1 \sin^{j-1}\bar{\phi}_1 - 
        \cos^{i+1}\bar{\phi}_2 \sin^{j-1}\bar{\phi}_2 +
        (j - 1) \mathcal{L}_{i, j-2}
    \bigg)
\end{align}  \end{linenomath}
for $i < 2, j \ge 2$ and
\begin{linenomath}  \begin{align}
    \label{eq:Lijupi}
    \mathcal{L}_{i,j} &=
    \frac{1}{i + j}
    \bigg(
        \cos^{i-1}\bar{\phi}_2 \sin^{j+1}\bar{\phi}_2 - 
        \cos^{i-1}\bar{\phi}_1 \sin^{j+1}\bar{\phi}_1 +
        (i - 1) \mathcal{L}_{i-2, j}
    \bigg)
\end{align}  \end{linenomath}
for all other terms. In practice, we compute $\mathcal{S}_{l, i, j}$ and $\mathcal{C}_{l, i, j}$ recursively and $\mathcal{M}_{l, i, j}$ as a matrix dot product, which makes evaluation of $p_n$ efficient.

The integral along the limb of the star is
\begin{linenomath}  \begin{align}
    t_n &=
        \int_{\xi_1}^{\xi_2}
        \x(\xi)^{\frac{l - m + 2}{2}}
        \left(\frac{\y(\xi)}{1-f'}\right)^{\frac{l + m}{2}}
        \frac{dy(\xi)}{d\xi} d\xi
        \nonumber\\
        &=
        (1-f')
        \mathcal{H}_{\frac{l - m + 4}{2}, \frac{l + m}{2}}
        \label{eq:tn_case1}
\end{align}  \end{linenomath}
where
\begin{linenomath}  \begin{align}
    \label{eq:Hlij}
    \mathcal{H}_{i,j} &\equiv
    \int_{\xi_1}^{\xi_2} \cos^i w \sin^j w \, d w
\end{align}  \end{linenomath}
is computed in the same way as Equation~(\ref{eq:Lij}).

\subsection{Case 2: $l = 1, m = 0$}
The remaining cases (2---5) have closed forms for $t_n$ but not $p_n$.
In case 2, the integral along the limb of the occultor is
\begin{linenomath}  \begin{align}
    p_n &=
        \frac{1}{3}
        \int_{\phi_1}^{\phi_2}
        \left(\frac{1-z'(\phi)^3}{1-z'(\phi)^2}\right)
        \left(-\y(\phi) \, \frac{dx(\phi)}{d\phi}  + \x(\phi) \, \frac{dy(\phi)}{d\phi}\right)
        d\phi
        \nonumber\\
        &=
        \frac{r}{3}
        \int_{\phi_1}^{\phi_2}
        (r + b \sin \phi)
        \left(\frac{1-z'(\phi)^3}{1-z'(\phi)^2}\right)
        d\phi
        \quad,
        \label{eq:pn_case2}
\end{align}  \end{linenomath}
where $z'(\phi)$ is given by Equation~(\ref{eq:zprime}). This integral is notoriously difficult to solve in the case of a spherical star, as it involves complicated functions of the elliptic integrals of the first, second, and third kinds (see, e.g., Equations 22--34 in \citealt{Pal2012}). In the case of an oblate star, it may not admit a closed form solution at all. We therefore solve this integral numerically by Gaussian quadrature; see below for details.

The integral along the limb of the star, on the other hand, reduces to a straightforward closed form:
\begin{linenomath}  \begin{align}
    t_n &=
        \frac{1}{3}
        \int_{\xi_1}^{\xi_2}
        \left(\frac{1-z'(\xi)^3}{1-z'(\xi)^2}\right)
        \left(-\y(\xi) \, \frac{dx(\xi)}{d\xi}  + \x(\xi) \, \frac{dy(\xi)}{d\xi}\right)
        d\xi
        \nonumber\\
        &=
        \frac{1}{3}
        (1 - f')
        (\xi_2 - \xi_1)
        \quad.
        \label{eq:tn_case2}
\end{align}  \end{linenomath}

\subsection{Case 3: $m = l - 1, l$ even}
In this case, the integral along the limb of the occultor is
\begin{linenomath}  \begin{align}
    p_n &=
        (1 - f')
        \int_{\phi_1}^{\phi_2}
        x(\phi)^{l-2}
        z'(\phi)^3
        \frac{dx(\phi)}{d\phi}
        d\phi
        \nonumber\\
        &=
        -r (1 - f')
        \int_{\phi_1}^{\phi_2}
        \big(
            r \cos\bar{\phi}
            + b \sin \theta
        \big)^{l-2}
        \sin\bar{\phi}
        z'(\phi)^3
        d\phi
        \quad,
        \label{eq:pn_case3}
\end{align}  \end{linenomath}
where we once again use  $\bar{\phi} \equiv \phi - \theta$. This integral is a ``pseudo-elliptic'' integral that reduces to a function of incomplete elliptic integrals of the first and second kinds when $f' = 0$. While a closed form solution may be possible in the oblate case, we were unable to find one. While it is possible to Taylor expand the integrand in the expression above about $f' = 0$ and obtain closed form solutions for each order individually, we find that this becomes computationally intractable above the quadratic term. Instead, as in the previous case, we opt to solve this integral numerically (details below).

On the other hand, the integral along the limb of the star is trivial:
\begin{linenomath}  \begin{align}
    t_n &=
        (1 - f')
        \int_{\xi_1}^{\xi_2}
        x(\xi)^{l-2}
        z'(\xi)^3
        \frac{dx(\xi)}{d\xi}
        d\xi
        \nonumber\\
        &=
        0
        \quad,
        \label{eq:tn_case3}
\end{align}  \end{linenomath}
since $z'(\xi)$ is zero everywhere along the limb of the star.

\subsection{Case 4: $m = l - 1, l$ odd}
The integral along the limb of the occultor is
\begin{linenomath}  \begin{align}
    p_n &=
        \int_{\phi_1}^{\phi_2}
        x(\phi)^{l-3}
        y(\phi)
        z'(\phi)^3
        \frac{dx(\phi)}{d\phi}
        d\phi
        \nonumber\\
        &=
        -r
        \int_{\phi_1}^{\phi_2}
        \big(
            r \cos\bar{\phi}
            + b \sin \theta
        \big)^{l-3}
        \big(
            r \sin \bar{\phi} +
            b \cos \theta
        \big)
        \sin\bar{\phi}
        z'(\phi)^3
        d\phi
        \quad,
        \label{eq:pn_case4}
\end{align}  \end{linenomath}
which we again solve numerically.
As before, the integral along the limb of the star is zero:
\begin{linenomath}  \begin{align}
    t_n &=
        (1 - f')
        \int_{\xi_1}^{\xi_2}
        x(\xi)^{l-3}
        y(\xi)
        z'(\xi)^3
        \frac{dx(\xi)}{d\xi}
        d\xi
        \nonumber\\
        &=
        0
        \quad.
        \label{eq:tn_case4}
\end{align}  \end{linenomath}

\subsection{Case 5: remaining terms}
The integral along the limb of the occultor for the final case is
\begin{linenomath}  \begin{align}
    p_n &=
        \int_{\phi_1}^{\phi_2}
        x(\phi)^\frac{l-m-3}{2}
        \left(\frac{y(\phi)}{1-f'}\right)^\frac{l+m-1}{2}
        z'(\phi)^3
        \frac{dy(\phi)}{d\phi}
        d\phi
        \nonumber\\
        &=
        r(1-f')^{-\frac{l+m-1}{2}}
        \int_{\phi_1}^{\phi_2}
         (r \cos \bar{\phi} + b \sin\theta)^\frac {l - m - 3} {2} (r \sin \bar{\phi} + b\cos\theta )^\frac {l + m - 1}{2}
         \cos \bar{\phi}
         z'(\phi)^3
        d\phi
        \quad,
        \label{eq:pn_case5}
\end{align}  \end{linenomath}
which, again, we solve numerically.
The integral along the limb of the star is once more zero:
\begin{linenomath}  \begin{align}
    t_n &=
        (1 - f')
        \int_{\xi_1}^{\xi_2}
        x(\xi)^\frac{l-m-3}{2}
        \left(\frac{y(\xi)}{1-f'}\right)^\frac{l+m-1}{2}
        z'(\xi)^3
        \frac{dy(\xi)}{d\xi}
        d\xi
        \nonumber\\
        &=
        0
        \quad.
        \label{eq:tn_case5}
\end{align}  \end{linenomath}

\subsection{Numerical integration}
Cases 2---5 in the computation of the $p_n$ integral do not admit straightforward closed form solutions. As we mentioned above, it is possible to derive approximate expressions to these integrals by linearizing the dependence on $f'$. While this allows one to derive efficient upward recursion relations in $n$, the approximation may not be sufficiently accurate for values of $f$ greater than a few percent. However, since both limb darkening and gravity darkening are well modeled by low order spherical harmonic expansions (see \S\ref{sec:analytic}), it is not in general necessary to compute the terms in the solution vector above $l_\mathrm{max}{\sim}6$. Moreover, cases 2--5 account for only about half of the terms in $\bvec{s_\obl^{\boldsymbol{\top}}}$, so in total fewer than 20 such integrations must be performed per point in the light curve. We find that a vectorized implementation of fixed Gauss-Legendre quadrature with 100 sample points for these terms is comparable in accuracy to and only marginally slower than the closed form solutions presented above. Within \starry, we therefore opt to integrate $p_n$ numerically in cases 2--5.

\begin{figure}[t!]
\begin{centering}
\includegraphics[width=0.75\linewidth]{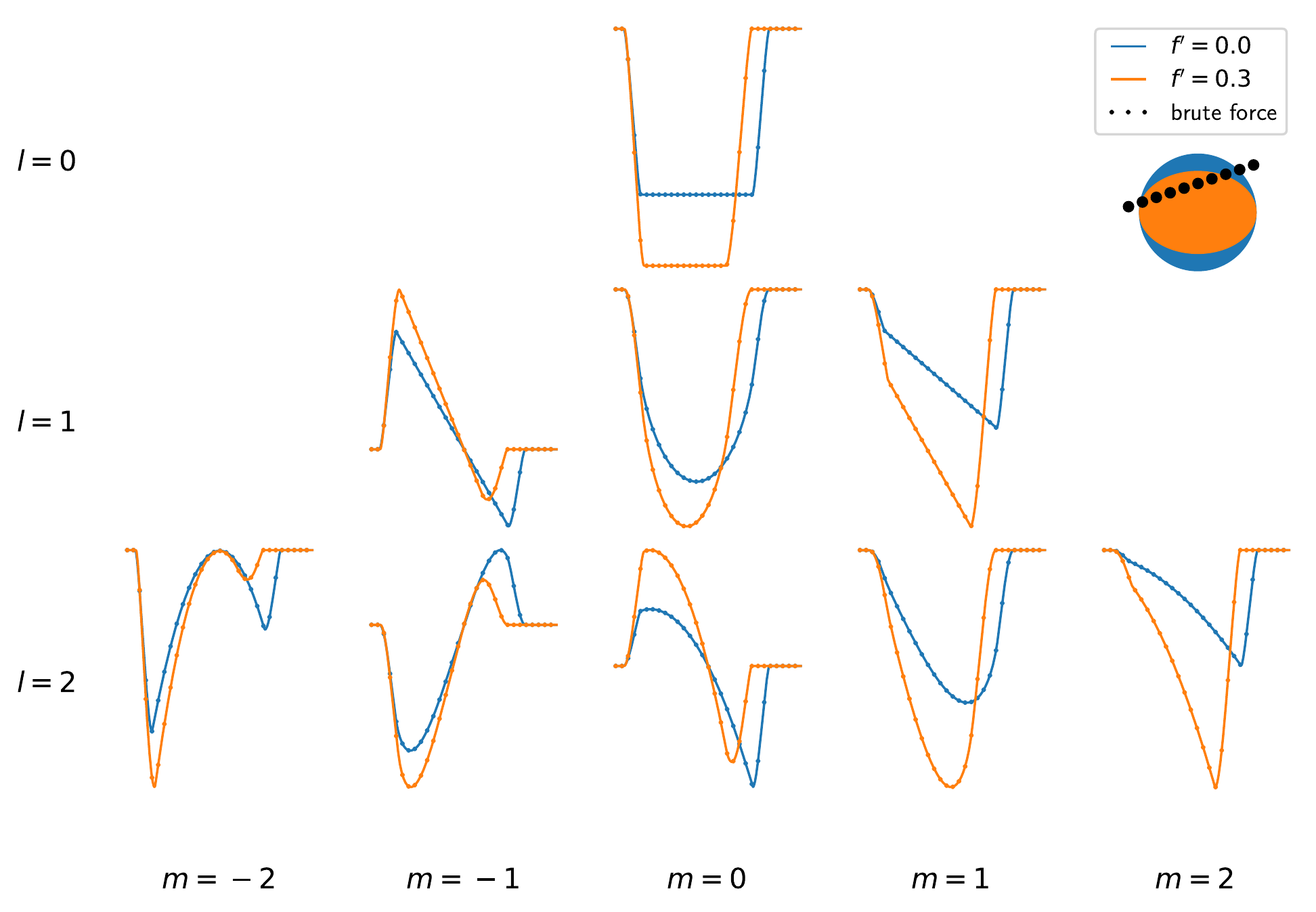}
\caption{%
The first nine terms in the solution vector $\bvec{s_\obl^{\boldsymbol{\top}}}$ over the course of an occultation by a planet of radius $r_\mathrm{o} = 0.1$ across a star with projected flattening $f' = 0$ (blue) and $f' = 0.3$ (orange); the inset at the top right shows the occultation geometry for both cases. Solid lines correspond to the solution derived in this paper, and dots to a brute force solution obtained by discretizing the stellar surface at high resolution and summing the intensities of the unocculted pixels. All light curves are normalized by the quantity $1 - f'$ so that the total unocculted flux is the same in both the spherical and oblate cases.
}
\label{fig:s_comparison}
\end{centering}
\end{figure}

Figure~\ref{fig:s_comparison} shows the first nine terms (up to $l_\mathrm{max} = 2$) of the solution vector over the course of an occultation by a planet of radius $r_\mathrm{o} = 0.1$. The curves for the spherical $(f' = 0)$ case are shown in blue and for a highly oblate $(f' = 0.3)$ case in orange; refer to the inset at the top right for the occultation geometry. Each curve corresponds to the transit light curve across each term in the Green's basis; the observed light curve is simply a linear combination of these basis light curves. The light curves in the oblate case have a shorter duration, given the shorter length of the transit chord across the ellipse; note that the larger depth in some of the panels is primarily due to the normalization. For reference, the blue and orange dots correspond to a brute force solution in which the integrals were computed by summing pixel intensities on a high resolution grid. In both cases, the portion of the star below the projected disk of the planet was gridded up into $10^6$ pixels, whose summed intensity was subtracted from the total intensity across the stellar disk. Our solution agrees with the numerical solution in all cases to within the discretization error of the latter.

\end{document}